\documentclass[11pt,a4paper]{article}

\usepackage{jcappub}
\usepackage{graphicx}
\usepackage{amssymb}
\usepackage{amsmath}
\usepackage{bm}
\usepackage{latexsym}
\usepackage{epsfig}
\usepackage{psfrag}
\usepackage{ulem}
\usepackage{textcomp}
\usepackage{color}
\usepackage{pstricks}
\usepackage[utf8]{inputenc}

\makeatletter
\gdef\@fpheader{}
\makeatother

\def\nn{\nonumber} 
\def\pa{{\partial}}
\def\f{\frac}
\def\l{\left}
\def\r{\right}
\def\d{{\rm d}}
\def\Mpl{M_{_{\rm Pl}}}
\def\beq{\begin{equation}}
\def\eeq{\end{equation}} 
\def\beqa{\begin{eqnarray}}
\def\eeqa{\end{eqnarray}} 

\def\bA{\bar A}
\def\cA{\mathcal A}

\def\psb{{\mathcal P}_{_{\rm B}}}

\def\vk{{\bm k}}
\def\kT{k_{_{\rm T}}}
\def\vka{{\bm k}_{1}}
\def\vkb{{\bm k}_{2}}
\def\vkc{{\bm k}_{3}}
\def\ka{k_{1}}
\def\kb{k_{2}}
\def\kc{k_{3}}

\def\Mp{M_{_{\rm Pl}}}
\def\cG{{\cal G}}
\def\cP{{\cal P}}

\def\ee{\eta_{\rm e}}

\def\a1{\alpha_{_1}}
\def\b1{\beta_{_1}}
\def\de1{\delta_{_1}}
\def\g1{\gamma_{_1}}
\def\bnl{b_{_{\rm NL}}}
\def\fnl{f_{_{\rm NL}}}

\def\cH{{\mathcal H}}

\newcommand{\viz}{\textit{viz.~}}
\newcommand{\ie}{\textit{i.e.~}}

\begin{document}
\title{Cross-correlations between scalar perturbations
and magnetic fields\\ in bouncing universes}
\author{Debika Chowdhury$^\dag$,}
\emailAdd{debika@physics.iitm.ac.in}
\affiliation{$^\dag$Department of Physics, Indian Institute of 
Technology Madras, Chennai~600036, India}
\author{L.~Sriramkumar$^\dag$,}
\emailAdd{sriram@physics.iitm.ac.in}
\author{Marc Kamionkowski$^\ddag$} 
\affiliation{$^\ddag$Department of Physics and Astronomy, Johns Hopkins University, 
3400 N. Charles Street, Baltimore, MD~21218, U.S.A.}
\emailAdd{kamion@jhu.edu}


\abstract
{Bouncing scenarios offer an alternative to the inflationary paradigm
for the generation of perturbations in the early universe.
Recently, there has been a surge in interest in examining the issue of 
primordial magnetogenesis in the context of bouncing universes.
As in the case of inflation, the conformal invariance of the electromagnetic 
action needs to be broken in bouncing scenarios in order to generate primordial 
magnetic fields which correspond to observed strengths today.
The non-minimal coupling, which typically depends on a scalar field that
possibly drives the homogeneous background, leads to a cross-correlation
at the level of the three-point function between the perturbation in 
the scalar field and the magnetic fields.
This has been studied in some detail in the context of inflation and,
specifically, it has been established that the three-point function
satisfies the so-called consistency relation in the squeezed limit.
In this work, we study the cross-correlation between the magnetic fields 
and the perturbation in an auxiliary scalar field in a certain class of 
bouncing scenarios.
We consider couplings that lead to scale invariant spectra of the magnetic 
field and evaluate the corresponding cross-correlation between the 
magnetic field and the perturbation in the scalar field.
We find that, when compared to de Sitter inflation, the dimensionless 
non-Gaussianity parameter that characterizes the amplitude of the 
cross-correlations proves to be considerably larger in bouncing 
scenarios.
We also show that the aforementioned consistency condition governing
the cross-correlation is violated in the bouncing models.
We discuss the implications of our results.}
\keywords{Primordial magnetic fields, inflation, bouncing universes,
non-Gaussianities}


\maketitle


\section{Introduction}


Magnetic fields permeate the universe over a wide range of scales.
In addition to the detection of magnetic fields in astrophysical systems 
such as stars and galaxies, recent observations also point towards the 
prevalence of these fields on cosmological scales, \viz in the large
scale structures~\cite{Grasso:2000wj,Widrow:2002ud,
Kandus:2010nw,Widrow:2011hs,Durrer:2013pga,Subramanian:2015lua} and even
in the intergalactic medium~\cite{Neronov:1900zz,Tavecchio:2010mk}.
In galaxies and clusters of galaxies, the strengths of the magnetic 
fields have been measured to be a few micro Gauss, while in the 
intergalactic medium, they are estimated to be of the order of 
$10^{-17}$ Gauss at $1$~Mpc~\cite{Neronov:1900zz,Tavecchio:2010mk,
Dermer:2010mm,Vovk:2011aa,Tavecchio:2010ja,Dolag:2010ni,Taylor:2011bn,
Takahashi:2011ac,Huan:2011kp,Finke:2015ona}.
In contrast, using CMB observations by Planck and POLARBEAR, 
the upper bound on the magnetic fields has been arrived at to be about a few nano 
Gauss~\cite{Ade:2015cva,Ade:2015cao}, which has also been corroborated by 
the upper limits obtained from the NRAO VLA Sky Survey~\cite{Pshirkov:2015tua}.
It is well known that certain astrophysical processes, particularly the 
dynamo mechanism, can, in the presence of a seed field, augment the 
strength of the magnetic fields in galaxies.
The origin of such progenitor fields is usually attributed to 
primordial processes.

\par

Inflation is presently the most widely accepted paradigm to explain the 
generation and evolution of perturbations in the early universe.
The issue of magnetogenesis, \viz the origin of magnetic fields, has 
been widely studied in various inflationary scenarios and, by means 
of the breaking of conformal invariance of the electromagnetic field, 
it has been possible to obtain scale invariant magnetic fields of the 
relevant strengths over the correlation scales of 
interest~\cite{Ratra:1991bn,Bamba:2003av,Bamba:2006ga,Demozzi:2009fu,Martin:2007ue,
Campanelli:2008kh,Kanno:2009ei,Subramanian:2009fu,Urban:2011bu,
Durrer:2010mq,Byrnes:2011aa,Jain:2012jy, Kahniashvili:2012vt,
Cheng:2014kga,Bamba:2014vda,Fujita:2015iga,Campanelli:2015jfa,
Fujita:2016qab,Tsagas:2016fax}.
Nevertheless, such models are known to be afflicted by the backreaction 
and strong coupling problems~\cite{Demozzi:2009fu,Ferreira:2013sqa,
Ferreira:2014hma}.
Consequently, it seems worthwhile to investigate the generation of 
magnetic fields in scenarios that provide a feasible alternative 
to the inflationary framework.
Among such alternatives, the most popular ones are the so-called bouncing 
models~\cite{Finelli:2001sr,Martin:2001ue,Tsujikawa:2002qc,Peter:2002cn,
Martin:2003sf,Martin:2003bp, Peter:2003rg, Martin:2004pm,Allen:2004vz,
Battefeld:2004mn,Creminelli:2004jg,Geshnizjani:2005hc,Abramo:2007mp,
Peter:2008qz,Falciano:2008gt,Cardoso:2008gz,Novello:2008ra,Lehners:2008vx,
Battefeld:2014uga,Brandenberger:2016vhg,Cai:2014bea}.
In these models, the universe goes through a period of contraction followed 
by an expanding phase, with a `bounce' connecting the two epochs.
Lately, there have been some efforts towards understanding the generation of 
magnetic fields in the bouncing scenarios~\cite{Salim:2006nw,Membiela:2013cea,
Sriramkumar:2015yza,Chowdhury:2016aet,Ben-Dayan:2016iks,Koley:2016jdw,
Ito:2016fqp}.

\par

In a recent work~\cite{Chowdhury:2016aet}, we had analytically illustrated 
how scale invariant magnetic fields can be produced in a certain sub-class 
of symmetric bouncing scenarios.
Using these analytical solutions, it would be interesting to examine the 
cross-correlations of these magnetic
fields with scalar perturbations present in the bouncing scenarios.
These correlations have been examined before in the context of 
inflation~\cite{Caldwell:2011ra,Motta:2012rn,Jain:2012ga,Jain:2012vm,
Kunze:2013hy}.
The magnitude of the non-Gaussianities generated through such correlations
has been estimated to be quite large for inflation, and the shape of the
non-Gaussianities peaks in the flattened limit, \ie when 
the wavenumber of the scalar perturbation is twice the 
wavenumber associated with the two modes of the magnetic field.
In this work, we study the cross-correlations of the magnetic fields produced
in bouncing universes with the perturbations in an auxiliary scalar field.
Ideally, it would be more appropriate to evaluate the 
cross-correlation between the primordial magnetic fields and the curvature 
perturbation.
However, as is well known, examining the behavior of the curvature perturbations
in bouncing models necessitates considerable modeling, often involving 
more than one field (see, for instance, Ref.~\cite{Raveendran:2017vfx}).
Therefore, it would be instructive to first investigate the behavior of 
the cross-correlation of the magnetic fields with the perturbation in an
auxiliary scalar field and explore their ramifications.

\par

One of the most important characteristics of three-point functions is 
their behavior in the squeezed limit of the wavenumbers involved, \ie 
when one of the wavenumbers is assumed to be much smaller than the 
other two.
In inflation, typically, the amplitude of the
mode with the longest wavelength and, 
therefore, the smallest wavenumber, freezes on super-Hubble scales.
Therefore, in the squeezed limit, the three-point function can be 
completely expressed in terms of the two-point function through a 
relation referred to as the consistency condition 
(see Refs.~\cite{Maldacena:2002vr,Creminelli:2004yq,Kundu:2013gha,
Sreenath:2014nka,Sreenath:2014nca,Dimastrogiovanni:2014ina,
Dimastrogiovanni:2015pla} in the context of three-point 
functions involving scalar and tensor perturbations, and 
Refs.~\cite{Jain:2012ga,Jain:2012vm} for cross-correlations between 
the scalar perturbation and the magnetic fields).
However, in the case of the bouncing models that we shall study, the 
amplitude of the scalar perturbations grows strongly as one approaches 
the bounce (in this context, see Refs.~\cite{Chowdhury:2015cma,
Raveendran:2017vfx}).
This suggests that the consistency relation may not hold in such scenarios
(for a similar effect in the case of the tensor bispectrum, see 
Ref.~\cite{Chowdhury:2015cma}).
Therefore, it is of utmost significance to examine whether the consistency 
relation, which holds true for inflationary magnetogenesis, would be valid 
in the bouncing models.

\par

This paper is organized as follows. 
In the following section, we shall describe a few essential aspects 
concerning the evolution of the electromagnetic field in the presence
of non-minimal coupling.
We shall also evaluate the power spectra of the magnetic fields that 
arise in inflationary as well as bouncing scenarios in the presence of 
a non-minimal, power law coupling that is often considered in this 
context.
In Sec.~\ref{sec:cc-sp-mf}, after briefly revisiting the calculation of 
the cross-correlations between the perturbation in the scalar field and 
the magnetic fields in the context of inflation, we shall evaluate the 
corresponding three-point function in the matter
bounce scenario of our interest.
In Sec.~\ref{sec:bnl}, we shall define a dimensionless non-Gaussianity 
parameter to characterize the three-point function and calculate the parameter 
in both the inflationary and bouncing models.
In Sec.~\ref{sec:sq-cons}, we shall evaluate the cross-correlation in the 
squeezed limit and illustrate that, while the consistency relation holds 
true in the case of inflation, it is violated in the bouncing scenario.
Finally, we shall conclude with a brief discussion in Sec.~\ref{sec:dis}.

\par

We shall work with natural units such that $\hbar=c=1$, and set the Planck 
mass to be $\Mpl=\l(8\,\pi\, G\r)^{-1/2}$. 
We shall adopt the metric signature of $(-, +, +, +)$. 
Greek indices shall denote the spacetime coordinates, whereas the Latin indices 
shall represent the spatial coordinates, except for $k$ which shall be reserved 
for denoting the wavenumber. 
Lastly, an overprime shall denote differentiation with respect to the conformal 
time coordinate.


\section{Generation of scale invariant magnetic fields in the early universe}
\label{sec:gen-mag-f}

In this section, we shall describe the generation of primordial magnetic
fields via a non-minimal coupling.
We shall introduce the form of the electromagnetic action and, after choosing 
to work in a specific gauge, we shall obtain the equation of motion governing 
the electromagnetic vector potential.
We shall quantize the vector potential in terms of the modes that satisfy 
the equation of motion in Fourier space.
We shall also define the power spectrum corresponding to the energy density 
associated with the magnetic field.
Thereafter, we shall consider a certain form of the coupling function wherein 
it can be expressed as a simple power of the scale factor, and analytically 
evaluate the magnetic power spectra in de Sitter inflation and in a specific 
class of bouncing scenarios.


\subsection{Non-minimally coupled electromagnetic fields}

We shall consider the background to be the spatially flat, 
Friedmann-Lema\^itre-Robertson-Walker (FLRW) metric that is described by 
the line-element
\begin{equation}
\d s^2 = a^2(\eta)\, 
\l(-\d\eta^2+\delta_{ij}\, \d x^i\,\d x^j\r),
\end{equation}
where $a(\eta)$ is the scale factor and $\eta$ denotes the conformal time 
coordinate.
We shall consider the action
\begin{equation}
S_{\rm em}[A^{\mu},\phi]
=-\frac{1}{16\,\pi} \int \d^{4}x\, \sqrt{-g}\,J^2(\phi)\, F_{\mu\nu}F^{\mu\nu},
\label{eq:em-action}
\end{equation}
where the electromagnetic field tensor $F_{\mu\nu}$ is given in terms of 
the vector potential $A^{\mu}$ by the relation
\begin{equation}
F_{\mu\nu} = \partial_\mu\,A_\nu - \partial_\nu\,A_\mu.
\end{equation}
The quantity $J(\phi)$ describes the non-minimal coupling, with $\phi$ 
denoting a scalar field that possibly contributes to the background 
evolution. 
We shall assume that there is no homogeneous component to the electromagnetic 
field.
We shall choose to work in the Coulomb gauge wherein $A_0 = 0$ and 
$\partial_i\,A^i = 0$.
In such a gauge, at the quadratic order in the inhomogeneous modes, the 
action describing the electromagnetic field is found to 
be (see, for example, Ref.~\cite{Martin:2007ue,Subramanian:2009fu})
\begin{equation}
S[A_i] 
= \frac{1}{4\,\pi}\,\int\,{\rm d}\,\eta\,\int\,{\rm d}^3\,\bm{x}
\left\{J^2\left(\phi\right)
\left[\frac{1}{2}\,A_i^{\prime\,2} - \frac{1}{4}\left(\partial_i\,A_j 
- \partial_j\,A_i\right)^2\right]\right\}.
\end{equation}
We can vary this action to arrive at the following equation of motion for 
the electromagnetic vector potential:
\begin{equation}
A_i^{\prime\prime} + 2\,\frac{J^\prime}{J}\,A_i^\prime - \nabla^2\,A_i 
= 0.
\end{equation}

\par

For each comoving wavevector $\bm{k}$, we can define the right-handed 
orthonormal basis $(\varepsilon_1^{\bm{k}}, \varepsilon_2^{\bm{k}}, 
\hat{\bm{k}})$, where
\begin{equation}
\vert\varepsilon^{\bm{k}}_i\vert^2 = 1,\quad
\varepsilon^{\bm{k}}_1\times\varepsilon^{\bm{k}}_2 
= \hat{\bm{k}}\quad{\rm and}\quad
\varepsilon^{\bm{k}}_1\cdot\varepsilon^{\bm{k}}_2
=\hat{\bm{k}}\cdot\varepsilon^{\bm{k}}_{1}  
= \hat{\bm{k}}\cdot\varepsilon^{\bm{k}}_{2}=0.
\end{equation}
On quantization, the vector potential $\hat{A_i}$ can be Fourier 
decomposed as follows~\cite{Martin:2007ue,Subramanian:2009fu}:
\begin{equation}
\hat{A_i}\l(\eta,{\bm x}\r) 
=\sqrt{4\,\pi}\int\frac{\d^3\,{\bm k}}{(2\,\pi)^{3/2}}\,
\sum_{\lambda=1}^2\,\varepsilon_{\lambda i}^{\bm k}\,
\l[\hat{b}_{\bm k}^{\lambda}\, {\bar A}_k(\eta)\,
{\rm e}^{i\,{\bm k}\cdot{\bm x}}
+ \hat{b}_{\bm k}^{\lambda\dagger}\, {\bar A}_k^{\ast}(\eta)\,
{\rm e}^{-i\,{\bm k}\cdot{\bm x}}\r],\label{eq:vp-d}
\end{equation}
where the Fourier modes ${\bar A}_{k}$ satisfy the differential equation
\begin{equation}
{\bar A}_k''+2\,\f{J'}{J}\,{\bar A}_k'+k^2\, {\bar A}_k=0.
\label{eq:de-Abk-o}
\end{equation}
The quantities $\varepsilon_{\lambda i}^{\bm k}$ represent the polarization 
vectors and the summation corresponds to the two orthonormal transverse 
polarizations. 
The operators $\hat{b}_{\bm k}^{\lambda}$ and 
$\hat{b}_{\bm k}^{\lambda\dagger}$ are the annihilation and creation 
operators satisfying the following standard commutation relations:
\begin{equation}
[\hat{b}_{\bm k}^{\lambda},\hat{b}_{\bm k'}^{\lambda'}] 
= [\hat{b}_{\bm k}^{\lambda\dagger},
\hat{b}_{\bm k'}^{\lambda'\dagger}] = 0,\quad
[\hat{b}_{\bm k}^{\lambda},
\hat{b}_{\bm k'}^{\lambda'\dagger}] 
=\delta^{(3)}\l({\bm k} - {\bm k'}\r)\,\delta_{\lambda\lambda'}.
\end{equation}
Let us now define a new variable ${\mathcal A}_k=J\, {\bar A}_k$, 
which, as we shall see, proves to be convenient to deal with.
In terms of the new variable, Eq.~(\ref{eq:de-Abk-o}) for ${\bar A}_k$ 
simplifies to 
\begin{equation}
{\mathcal A}_k''+\l(k^2-\f{J''}{J}\r)\,{\mathcal A}_k=0.\label{eq:de-cAk}
\end{equation}


\subsection{Power spectra}

Since we shall be focusing only on the properties of the magnetic field 
and not those of the electric field, let us attend to the power spectra 
of the generated magnetic fields.
Let $\hat{\rho}_{_{\rm B}}$ denote the operator corresponding to the 
energy density associated with the magnetic field.
Upon using the decomposition~(\ref{eq:vp-d}) of the vector potential, the 
expectation values of the energy density $\hat{\rho}_{_{\rm B}}$ can be 
evaluated in the vacuum state, say, $\vert 0\rangle$, that is annihilated 
by the operator ${\hat b}_{\bm k}^{\lambda}$ for all ${\bm k}$ and $\lambda$.
It can be shown that the spectral energy density of the magnetic field can 
be expressed in terms of the modes ${\bar A}_k$ and $\cA_k$ and the coupling 
function $J$ as follows~\cite{Martin:2007ue,Subramanian:2009fu}:
\begin{eqnarray}
\psb(k)
=\frac{\d\langle 0 \vert \hat{\rho}_{_{\rm B}}\vert 0 \rangle}
{\d\,{\rm ln}\,k}
=\f{J^2(\eta)}{2\,\pi^{2}}\,\f{k^{5}}{a^{4}(\eta)}\,
\vert{\bar A}_k(\eta)\vert^{2}
=\f{1}{2\,\pi^{2}}\,\f{k^{5}}{a^{4}(\eta)}\,
\vert \cA_k(\eta)\vert^{2}.\label{eq:psb}
\end{eqnarray}
The spectral energy density $\psb(k)$ is often referred to as the power 
spectrum of the generated magnetic fields. 
A flat or scale invariant magnetic field spectrum corresponds to a 
constant, \ie $k$-independent, ${\mathcal P}_{_{\rm B}}(k)$.


\subsection{The inflationary case}

Let us first consider the simple case of de Sitter inflation, wherein the 
scale factor is given by
\begin{equation}
a(\eta) = -\frac{1}{H_0\,\eta},\label{eq:sf-ds}
\end{equation}
with $H_0$ being the value of the Hubble parameter during inflation.
In order to solve for the electromagnetic modes, we need to choose a form 
of the coupling function.
In keeping with the expressions for the coupling functions that have been 
adopted earlier~\cite{Martin:2007ue,Subramanian:2009fu,Caldwell:2011ra,
Motta:2012rn,Jain:2012ga,Jain:2012vm}, we shall work with a coupling function 
that can be written as a simple power of the scale factor as follows:
\begin{equation}
J(\eta) = J_0\,[a(\eta)]^n
=\f{J_0}{(-H_0\,\eta)^n}.\label{eq:Ji}
\end{equation}
We shall set $J_0=(-H_0\,\ee)^n$, where $\ee$ denotes the conformal time at
the end of inflation. 
This choice ensures that $J$ reduces to unity at $\ee$.
For such a coupling function, $J''/J=n\,(n+1)/\eta^2$ and the solutions to 
Eq.~(\ref{eq:de-cAk}) can then be expressed 
as~\cite{Martin:2007ue,Subramanian:2015lua}:
\begin{equation}\label{eq:cA-infl-d1}
\cA_k(\eta)  
= \sqrt{-k\, \eta}\, \l[C_{1}(k)\, J_{n+1/2}(-k\, \eta) 
+ C_{2}(k)\, J_{-n-1/2} (-k\,\eta)\r],
\end{equation}
where $J_\nu(x)$ refers to the Bessel function and the coefficients 
$C_1(k)$ and $C_2(k)$ can be fixed from the standard Bunch-Davies 
initial conditions to be
\begin{eqnarray}
C_1(k)=\sqrt{\f{\pi}{4\,k}}\, 
\f{{\rm e}^{-i\,(n-1)\,\pi/2}}{{\rm cos}(n\pi)}, \qquad
C_2(k)=\sqrt{\f{\pi}{4\,k}}\, 
\f{{\rm e}^{i\,n\,\pi/2}}{{\rm cos}(n\pi)}.
\end{eqnarray}
We should emphasize here that the initial conditions are 
imposed in the domain wherein $k^2 \gg J''/J$, which, for the simple cases 
of de Sitter inflation and a power law coupling [cf.~Eq.~\ref{eq:Ji}], 
roughly corresponds to the modes being well inside the Hubble radius.
It is useful to note that the mode ${\bar A}_k$ can be expressed 
conveniently in terms of the Hankel function of the first kind
$H_\nu^{(1)}(x)$ as
\begin{equation}
\bA_k(\eta) 
= \f{1}{J(\eta)}\,\sqrt{-\f{\pi\,\eta}{4}}\,
{\rm e}^{i\,(n+1)\,\pi/2}\,H_{n+1/2}^{(1)}(-k\eta),\label{eq:Abki}
\end{equation}
with $J(\eta)$ given by Eq.~(\ref{eq:Ji}).

\par

Using the above expression for the modes and the definition~(\ref{eq:psb}),
the power spectrum for the magnetic field can be obtained to be
\begin{equation}
\psb(k) = \frac{H_0^4}{8\,\pi}\,(-k\,\ee)^5\,
\vert H_{n+1/2}^{(1)}(-k\,\ee)\vert^2,
\end{equation}
For $n>-1/2$, at late times (\ie as $\ee\to 0$), this expression reduces to
\begin{equation}
\psb(k) = \frac{H_0^4}{8\,\pi^3}\,
\l[2^{2\,n+1}\,\Gamma(\tfrac{1}{2}+n)^2\r]\,
(-k\,\ee)^{4-2\,n}.
\label{eq:psbi}
\end{equation}
Therefore, the spectral index characterizing the power spectrum of the 
magnetic field is given by $n_{_{\rm B}} = 4-2\,n$.
Clearly, we must have $n=2$ in order to obtain a scale invariant power 
spectrum.
On the other hand, for $n<-1/2$, at late times 
(\ie as $\ee\to 0$), the power spectrum can be obtained to be
\begin{equation}
\psb(k) = \frac{H_0^4}{8\,\pi^3}\,
\l[2^{-2\,n-1}\,\Gamma(-\tfrac{1}{2}-n)^2\r]\,
(-k\,\ee)^{6+2\,n}.
\end{equation}
In this case, the spectral index of the magnetic field power spectrum
is given by $n_{_{\rm B}} = 6+2\,n$.
Therefore, a scale invariant power spectrum can also be
obtained for $n=-3$.
However, the cases wherein $n<0$ lead to severe backreaction issues which
result in the termination of inflation within about $10$ 
e-folds~\cite{Caldwell:2011ra,Jain:2012vm}.
Therefore, we shall avoid such scenarios in this work.
Also, note that, when $n_{_{\rm B}}=0$, the amplitude 
of the power spectrum is determined only by the value of $H_0$.
One finds that, in order to obtain magnetic fields of nano Gauss strength 
today, the value of $H_0$ should be of the order of 
$10^{13}\,{\rm GeV}$~\cite{Caldwell:2011ra}.


\subsection{The bouncing scenarios}\label{subsec:psb-bu}

We shall now discuss the generation of magnetic fields in bouncing 
scenarios.
Let us assume that the non-singular bouncing scenario of our interest 
is described by the following form of the scale factor 
$a(\eta)$ (in this context, see Refs.~\cite{Sriramkumar:2015yza,
Chowdhury:2015cma,Raveendran:2017vfx}):
\begin{equation}
a(\eta)=a_{0}\, \l(1+\f{\eta^2}{\eta_{0}^{2}}\r)^{q}
= a_{0}\, \l(1+k_0^2\,\eta^2\r)^{q},\label{eq:sf}
\end{equation}
where $a_{0}$ is the minimum value of the scale factor attained at the 
bounce (\ie at $\eta=0$), $\eta_{0}=1/k_0$ denotes the time scale of 
the duration of the bounce and $q>0$. 
It is evident that, for the case $q=1$, during very early times 
wherein $\eta \ll -\eta_0$, the scale factor behaves as $a \propto 
\eta^2$, which corresponds to the behavior of the scale factor in a 
matter dominated universe. 
Therefore, the $q=1$ case is often referred to as the matter bounce 
scenario.
Later, when calculating the three-point function of interest, we shall 
restrict ourselves to the case of the matter bounce, but for now let us 
consider an arbitrary $q$.
Evidently, we shall also be required to assume a form of the non-minimal 
coupling function $J$.
As in the case of inflation, we shall assume that the coupling function 
can be expressed as a power law of the scale factor as follows:
\begin{equation}
J(\eta)=J_0\, \l[a(\eta)\r]^{\bar n}
=J_0\,a_{0}^{\bar n}\, \l(1+k_0^2\,\eta^2\r)^{{\bar n}\,q},\label{eq:Jb}
\end{equation}
where $J_0$ is a suitable constant.
We shall comment on the choice of $J_0$ in due course.

\par

In order to solve the equation of motion governing the evolution of the
electromagnetic mode in the matter bounce scenario, for convenience, we 
shall divide the time period of our interest into two domains, one far 
away from and prior to the bounce (\ie $-\infty<\eta<-\alpha\,\eta_0$,
with $\alpha\gg 1$) and another around the bounce (\ie $-\alpha\,\eta_0 
<\eta <\beta\,\eta_0$, with $\beta>0$).
We shall assume that $\alpha$ is of the order of $10^5$ and, as we shall
explain later, we shall set $\beta$ to be $10^2$. 
During the first domain, the non-minimal coupling function $J$ simplifies
to a power law form:~$J(\eta) \propto \eta^{\bar \gamma}$, 
where ${\bar \gamma} =2\,{\bar n}\,q$.
Under these conditions, we have $J''/J\simeq {\bar \gamma}\,({\bar \gamma}-1)/\eta^2$,
which is of the same form as in the inflationary case that we had discussed 
in the previous sub-section.
Therefore, in the first domain, we can express the electromagnetic modes 
$\bA_k$ in terms of the Hankel function $H_\nu^{(1)}(x)$ as 
follows~\cite{Sriramkumar:2015yza,Chowdhury:2016aet}:
\begin{equation}
\bA_k(\eta)  
\simeq \f{1}{J(\eta)}\,\sqrt{-\f{\pi\,\eta}{4}}\;
{\rm e}^{i\,{\bar \gamma}\,\pi/2}\,H_{{\bar \gamma}-1/2}^{(1)}(-k\eta),\label{eq:Ab-d1}
\end{equation}
with $J(\eta)\simeq J_0\,a_0^{\bar n}\,(k_0\,\eta)^{\bar \gamma}$.
We should emphasize again that this solution has been constructed 
assuming that the Bunch-Davies initial conditions are imposed on the
modes when $k^2 \gg J''/J$.
As we have clarified in the context of inflation, for the form of the
coupling function we are working with [cf.~Eq.~(\ref{eq:Jb})], this
condition again corresponds to the modes of interest being well inside
the Hubble radius.
\par

Let us now evaluate the power spectra of the magnetic fields as one 
approaches the bounce, in fact as $k\,\eta \to 0^-$.
Since the solution~(\ref{eq:Ab-d1}) is assumed to be applicable only 
in the first domain (\ie over $-\infty<\eta<-\alpha\,\eta_0$), we need to 
restrict ourselves to wavenumbers such that $k\ll k_0/\alpha$. 
The power spectra of the magnetic fields for such modes can be written 
as~\cite{Sriramkumar:2015yza,Chowdhury:2016aet}
\begin{equation}
{\cal P}_{_{\rm B}}(k) 
\simeq \frac{{\cal F}(m)}{2\,\pi^2}\, \l(\f{H}{2\, q}\r)^4 (-k\,\eta)^{4+2\,m},
\label{eq:psb-bb}
\end{equation}
where $H\simeq (2\,q/a_0\,\eta)\, (\eta_0/\eta)^{2\,q}$, while $m={\bar \gamma}$ for 
${\bar \gamma} \le 1/2$ and $m =1-{\bar \gamma}$ for ${\bar \gamma} \ge 1/2$.
The quantity ${\cal F}(m)$ is given by
\begin{equation}
{\cal F}(m) = \frac{\pi}{2^{2\,m+1}\,\Gamma^2(m+1/2)\,\cos^2(\pi\,m)}.
\end{equation}
Evidently the value of the index $m=-2$, which corresponds to either ${\bar \gamma}=3$ 
or ${\bar \gamma}=-2$, leads to a scale invariant spectrum for the magnetic field.

\par

The power spectrum~(\ref{eq:psb-bb}) has been calculated before the bounce.
We need to evolve the modes across the bounce and evaluate the power spectrum
of the magnetic field {\it after}\/ the bounce.
In order for the problem to be tractable analytically, we shall hereafter 
restrict our analysis to the cases wherein ${\bar n}>0$~\cite{Chowdhury:2016aet}.
We find that for scales of cosmological interest such that $k\ll k_0$, 
$k^2\ll J''/J$ around the bounce.
Hence, near the bounce, we can neglect the $k^2$ term in
Eq.~(\ref{eq:de-Abk-o}) to arrive at
\begin{equation}
\bA_k''+2\,\f{J'}{J}\, \bA_k'\simeq 0.
\end{equation}
This equation can be integrated twice to yield~\cite{Chowdhury:2016aet}
\begin{eqnarray}
\bA_k(\eta)
\simeq \bA_k(\eta_\ast)+\bA_k'(\eta_\ast)\,\f{a^{2\,{\bar n}}(\eta_\ast)}{a_0^{2\,{\bar n}}}
\biggl[\eta\;{}_2F_1\l(\f{1}{2},{\bar \gamma};\f{3}{2};
-\f{\eta^2}{\eta_0^2}\r)
-\eta_\ast\;{}_2F_1\l(\f{1}{2},{\bar \gamma};\f{3}{2};
-\f{\eta_\ast^2}{\eta_0^2}\r)\biggl],\nn\label{eq:Ab-d2}\\
\end{eqnarray}
where ${}_2F_1(a,b;c;z)$ denotes the hypergeometric function (see, for instance, 
Ref.~\cite{Mathematica11.0}).
We shall choose $\eta_\ast=-\alpha\,\eta_0$, which then permits us to use the 
solution~(\ref{eq:Ab-d1}) in first domain to determine $\bA_k(\eta_\ast)$ and
$\bA_k'(\eta_\ast)$, and thereby arrive at the solution in the second domain.

\par

Note that the solution~(\ref{eq:Ab-d2}) is valid only when $k^2\ll J''/J$
and, evidently, the condition will fail at suitably late times after the 
bounce (just as it does prior to the bounce).
Therefore, the solution has to be utilized to evaluate the power spectrum of
the magnetic field in the domain of its validity.
Moreover, the numerical analysis of the modes suggest that the analytical 
solutions that we have obtained would be invalidated well before the 
condition $k^2= J''/J$ is satisfied after the bounce~\cite{Sriramkumar:2015yza}.
For these reasons, we shall choose to evaluate the power spectrum after 
the bounce at $\eta=\beta\,\eta_0$, with $\beta$ chosen to be about $10^2$.
Interestingly, for the scale factor~(\ref{eq:sf}) we are working with, for,
say, $q=1$, the time $\eta=\beta\,\eta_0$ roughly corresponds to the time
of reheating that follows the phase of inflation in the conventional hot
big bang model~\cite{Sriramkumar:2015yza}.
Having explained the reasons behind our choice of $\beta$, 
this seems an ideal stage to mention the value of $J_0$ we shall work with.
It is straightforward to show that the power spectrum of the magnetic field 
does not depend on~$J_0$.
However, as we shall see later, the three-point cross-correlation indeed 
depends on its value.
We shall choose the value of $J_0$ such that $J(\beta\,\eta_0)=1$.
This ensures that we recover the standard electromagnetic coupling after the 
bounce at roughly the time when the universe is expected to transit to the 
radiation dominated era, just as is done in the context of inflation.
We can now analytically evaluate the power spectrum after the bounce at $\eta
=\beta\,\eta_0$.
We find that the power spectrum of the magnetic field retains its 
scale dependence across the bounce~\cite{Chowdhury:2016aet}.
In the scale invariant case corresponding to ${\bar \gamma}=3$, the amplitude 
of the power spectrum for $k/(\alpha\,k_0)\ll 1$  can be determined to be
\begin{equation}
\psb(k) \simeq \l(\f{45}{16\,\beta}\r)^2\,\l(\f{k_0}{a_0}\r)^4.
\end{equation}
For instance, if we choose, $k_0/(a_0\,\Mpl)\simeq 10^{-4}$, we find that 
the above spectrum will lead to magnetic fields of nano Gauss strengths
today.


\section{Cross-correlations between the scalar perturbation and magnetic 
field}\label{sec:cc-sp-mf}

As is evident from the results of the preceding section, it is possible to 
obtain scale invariant magnetic fields of the requisite strength both in 
the case of inflation as well as in bouncing scenarios.
Therefore, the behavior of the two-point function of the primordial magnetic 
fields alone is not adequate to distinguish between the inflationary and 
bouncing scenarios.
It would hence be of utmost importance to study the cross-correlations of 
these fields with other fields that are expected to exist in the early 
universe, particularly the scalar fields.
In this section, we shall arrive at the expression for the three-point 
function involving the magnetic field and the perturbation in the scalar 
field which leads to the non-minimal coupling.
We shall first revisit the case of de Sitter inflation, wherein we shall 
consider the perturbation in an auxiliary scalar field and evaluate the 
three-point function.
Thereafter, we shall calculate the three-point function in the bouncing 
model of our interest.
We shall analyze the three-point function in these scenarios for two 
cases~--~one which leads to a scale invariant power spectrum for the 
magnetic field and another which results in a blue-tilted power spectrum 
that scales as the square of the wavenumber involved.
While the former is observationally relevant, the latter case involves
simpler computations and we shall utilize it to illustrate certain points.

\par 

The Lagrangian at the third order involving the perturbation $\delta \phi$ 
in the scalar field and the electromagnetic vector potential $A_i$ can be
easily obtained from the original action~(\ref{eq:em-action}).
The corresponding interaction Hamiltonian can be determined to
be~\cite{Caldwell:2011ra}
\begin{equation}
H_{\rm int} 
= \f{1}{4\,\pi}\,\int\,{\rm d}^3\,\bm{x}\,
J\,\f{{\rm d}J}{{\rm d}\phi}\,\f{\delta\phi}{\Mpl}\,
\l[A_i^{\prime\,2} + \f{1}{2}\l(\pa_i\,A_j - \pa_j\,A_i\right)^2\r], 
\label{eq:Hint}
\end{equation}
where $\delta\phi$ denotes the perturbation in the scalar field.
The cross-correlation between the perturbation in the scalar field and the 
magnetic field in real space is defined as
\begin{eqnarray}
& &\!\!\!\!\!\!\!\!\!\!\!\!\!\!\!\!\!\!\!\!\!\!\!\!
\l\langle\frac{\hat{\delta\phi}(\eta,\bm{x})}{\Mpl}\,
\hat{B}^i(\eta,\bm{x})\,\hat{B}_i(\eta,\bm{x})\r\rangle\nn\\ 
&=& \int \f{\d^3{\bm k}_1}{(2\,\pi)^{3/2}}
\int \f{\d^3{\bm k}_2}{(2\,\pi)^{3/2}}
\int \f{\d^3{\bm k}_3}{(2\,\pi)^{3/2}}\,
\l\langle \f{\hat{\delta\phi}_{\bm{k_1}}\!(\eta)}{\Mpl}\,
\hat{B}^i_{\bm{k_2}}(\eta)\,\hat{B}_{i\,\bm{k_3}}(\eta)\r\rangle\,
{\rm e}^{i\,({\bm k_1}+{\bm k_2}+{\bm k_3})\cdot{\bm x}},
\end{eqnarray}
where the components $B_i$ of the magnetic field are related to the 
vector potential $A_i$ through the relation
\begin{equation}
B_i = \f{1}{a}\,\epsilon_{ijl}\,\partial_j\,A_l,
\end{equation}
while $\delta\phi_{\bm k}$ and  $B^i_{\bm k}$ denote the Fourier modes 
associated with the perturbation in the scalar field and the $i$-th 
component of the magnetic field.
According to the standard rules of perturbative quantum field theory, the 
cross-correlation between the perturbation in the scalar field and the 
magnetic field in Fourier space, evaluated at the end of inflation, is 
given by~\cite{Caldwell:2011ra,Jain:2012vm}
\begin{equation}
\l\langle \frac{{\hat{\delta\phi}_{\bm{k_1}}\!(\ee)}}{\Mpl}\,
\hat{B}^i_{\bm{k_2}}(\eta_{\rm e})\,
\hat{B}_{i\,\bm{k_3}}(\eta_{\rm e})\r\rangle
= -i\, \int_{\eta_{\rm i}}^{\eta_{\rm e}} {\rm d}\eta\,
\l\langle\l[\frac{\hat{\delta\phi}_{\bm{k_1}}}{\Mpl}(\eta_{\rm e})\,
\hat{B}^i_{\bm{k_2}}(\eta_{\rm e})
\,\hat{B}_{i\,\bm{k_3}}(\eta_{\rm e}),\hat{H}_{\rm int}(\eta)\r]\r\rangle,
\label{eq:tpffn}
\end{equation}
where ${\hat H}_{\rm int}$ is the operator associated with the 
Hamiltonian~(\ref{eq:Hint}) and the square brackets indicates 
the commutator.

\par

We have already discussed the quantization of the electromagnetic modes
in the previous section. 
The perturbation in the scalar field can be quantized in terms of the 
corresponding Fourier modes, say, $f_k$ as 
\begin{equation}
\hat{\delta\phi}\l(\eta,{\bm x}\r) 
=\int\frac{\d^3\,{\bm k}}{(2\,\pi)^{3/2}}\,
\l[\hat{a}_{\bm k}\, f_k(\eta)\,{\rm e}^{i\,{\bm k}\cdot{\bm x}}
+ \hat{a}_{\bm k}^{\dagger}\, f_k^{\ast}(\eta)\,
{\rm e}^{-i\,{\bm k}\cdot{\bm x}}\r],
\end{equation}
where the annihilation and creation operators $\hat{a}_{\bm k}$ and 
$\hat{a}_{\bm k}^\dag$ satisfy the following standard commutation
relations:
\begin{equation}
[\hat{a}_{\bm k},\hat{a}_{\bm k'}] 
= [\hat{a}_{\bm k}^{\dagger},\hat{a}_{\bm k'}^{\dagger}] = 0,\quad
[\hat{a}_{\bm k},\hat{a}_{\bm k'}^{\dagger}] 
=\delta^{(3)}({\bm k} - {\bm k'}).\label{eq:cc-od}
\end{equation}
As we shall see, in order to achieve the coupling functions~(\ref{eq:Ji})
and~(\ref{eq:Jb}), we shall assume that the canonical scalar field is 
governed by the linear potential in the case of inflation and is free in
the case of the bounce.
We should emphasize that the field $\phi$ is actually an 
auxiliary scalar field that does not necessarily source the background.
In both the cases of inflation and bounce, the perturbation in the scalar field $\delta\phi$ 
is governed by the equation of motion
\begin{equation}
\delta\phi''+2\,{\cal H}\,\delta\phi'-\nabla^2\delta\phi=0,
\end{equation}
where ${\cal H}=a'/a$ is the conformal Hubble parameter.
The Fourier modes $f_k$ therefore satisfy the differential equation
\begin{equation}
f_k'' + 2\,\cH\,f_k' + k^2\,f_k=0.\label{eq:de-fk}
\end{equation}

\par

Let us now define
\begin{eqnarray}
\l\langle \f{\hat{\delta\phi}_{\bm{k_1}}(\ee)}{\Mpl}
\hat{B}^i_{\bm{k_2}}(\ee)\,
\hat{B}_{i\,\bm{k_3}}(\ee)\r\rangle
\equiv (2\,\pi)^{-3/2}\,
G_{\delta\phi B B}\l(\bm{k_1},\bm{k_2},\bm{k_3}\r)\,
\delta^{(3)}\l(\bm{k_1}+\bm{k_2}+\bm{k_3}\r).\nn\\
\label{eq:cc}
\end{eqnarray}
Then, upon using the expression~(\ref{eq:tpffn}), along with the form
of the interaction Hamiltonian~(\ref{eq:Hint}) and Wick's theorem that 
applies to the products of the operators ${\hat A}_{\bm k}^i$ and 
$\hat{\delta\phi}_{\bm k}$, one can show that the quantity 
$G_{\delta\phi B B}\l(\bm{k_1},\bm{k_2},\bm{k_3}\r)$ can be expressed as
\begin{eqnarray}
G_{\delta\phi B B}\l(\bm{k_1},\bm{k_2},\bm{k_3}\r)
& = &\frac{8\,\pi}{\Mp\,a^2(\ee)}\,f_{\ka}(\ee)\,
\bA_{\kb}(\ee)\,\bA_{\kc}(\ee)\,
\biggl\{2\left(\vkb\cdot\vkc\right)\,\cG_1\l(\bm{k_1},\bm{k_2},\bm{k_3}\r)\nn\\
& & -\, \left[\frac{\left(\vkb\cdot\vkc\right)^2}{k_2\,k_3} + k_2\,k_3\right]\,
\cG_2\l(\bm{k_1},\bm{k_2},\bm{k_3}\r)\biggr\} \nn\\
& & +\,{\rm complex~conjugate},
\end{eqnarray}
where $\cG_1\l(\bm{k_1},\bm{k_2},\bm{k_3}\r)$ and  
$\cG_2\l(\bm{k_1},\bm{k_2},\bm{k_3}\r)$ are described by the integrals
\begin{subequations}\label{eq:int-eta}
\begin{eqnarray}
\cG_1\l(\bm{k_1},\bm{k_2},\bm{k_3}\r) 
& =& i\,\int_{\eta_{\rm i}}^{\eta_{\rm e}} {\rm d}\eta\,
J\,\frac{{\rm d}J}{{\rm d}\phi}\,f_{\ka}^\ast(\eta)\,
{\bar A}_{\kb}^{\prime\ast}(\eta)\,
{\bar A}_{\kc}^{\prime\ast}(\eta),\label{eq:int-1-eta}\\
\cG_2\l(\bm{k_1},\bm{k_2},\bm{k_3}\r) 
& =& i\,k_2\,k_3\,\int_{\eta_{\rm i}}^{\eta_{\rm e}} {\rm d}\eta\,
J\,\frac{{\rm d}J}{{\rm d}\phi}\,f_{\ka}^\ast(\eta)\,
{\bar A}_{\kb}^{\ast}(\eta)\,{\bar A}_{\kc}^{\ast}(\eta).
\label{eq:int-2-eta}
\end{eqnarray}
\end{subequations}
Given the solutions for the electromagnetic modes $\bA_k$ and the 
scalar perturbations $f_k$ [cf. Eqs.~(\ref{eq:de-Abk-o}) 
and~(\ref{eq:de-fk})], the above integrals can be evaluated in the 
inflationary and bouncing scenarios to arrive at the three-point 
function $G_{\delta\phi B B}\l(\bm{k_1},\bm{k_2},\bm{k_3}\r)$.


\subsection{The three-point function in de Sitter inflation}

Before we go on to evaluate the three-point function in the bouncing scenario,
we shall first revisit its calculation in inflation in order to illustrate 
a few points. 
In Sec.~\ref{sec:gen-mag-f}, when we had considered the evolution of the 
electromagnetic modes, for simplicity, we had assumed that the non-minimal 
couping $J(\eta)$ is given by Eq.~(\ref{eq:Ji}).
In contrast, to evaluate the three-point function of interest, apart from 
$J$, we also need the behavior of $\d J/\d \phi$ [cf. Eqs.~(\ref{eq:int-eta})], 
which requires $J(\phi)$.
This can be arrived at easily.
Let the auxiliary scalar field $\phi$ that is evolving in de Sitter 
spacetime characterized by the scale factor~(\ref{eq:sf-ds}) be 
described by the potential $V(\phi)$.
In de Sitter spacetime, the homogeneous scalar field $\phi$ satisfies 
the following equation of motion:
\begin{equation}
\phi'' - \f{2}{\eta}\,\phi' + a^2\,V_\phi= 0,
\end{equation}
where $V_\phi=\d V/\d\phi$.
If we now assume that $V(\phi) = -3\,n\,\Mp\,H_0^2\,\phi$, where $n$ is
a constant, then it is straightforward to show that, for a suitable choice 
of initial conditions, the solution to the above equation governing the
scalar field can be written as~\cite{Caldwell:2011ra}
\begin{equation}
\phi(\eta) 
=-n\,\Mp\, {\rm ln}\, \eta.
\end{equation}
Therefore, upon setting $J(\phi)=J_0\, \exp\,(\phi/\Mpl)$, we can arrive 
at the desired behavior of $J(\eta)$ [as given by Eq.~(\ref{eq:Ji})]
that we had worked with.  
With $J(\phi)$ at hand, we can, evidently, obtain ${\rm d}J/{\rm d}\phi$
to be
\begin{eqnarray}
\frac{{\rm d}J}{{\rm d}\phi} &=& \frac{J(\phi)}{\Mp},
\label{eq:dJdphi-infl}
\end{eqnarray}
thereby arriving at the required quantities related to the background.

\par

We shall now evaluate the three-point function~$G_{\delta\phi B B}(\bm{k_1},
\bm{k_2},\bm{k_3})$ for two specific values of $n$, as it proves to be
difficult to evaluate the quantity for arbitrary $n$.
Therefore, we shall consider the two cases wherein $n=1$ and $n=2$.
The $n=1$ case leads to a blue-tilted power spectrum for the magnetic
field with the spectral index $n_{_{\rm B}} = 2$. 
Whereas, the $n=2$ case leads to the desired scale invariant spectrum.
Note that the behavior of the mode $f_k$ depends only on the scale 
factor [cf. Eq.~(\ref{eq:de-fk})]. 
As is well known, in de Sitter spacetime, the mode $f_k$ satisfying 
the standard Bunch-Davies initial condition is given by
\begin{equation}
f_k(\eta) = \frac{i\,H_0}{\sqrt{2\,k^3}}\,
\l(1 + i\,k\,\eta\right)\,{\rm e}^{-i\,k\,\eta}.\label{eq:fk-ds}
\end{equation}


\subsubsection{The case of $n=1$}

When $n=1$, the electromagnetic mode $\bA_k$ and its derivative $\bA_k'$ 
can be written as
\begin{subequations}
\begin{eqnarray}
\bA_k(\eta) 
&=& \sqrt{-\frac{\pi\,\eta}{4}}\,\l(\f{\eta}{\ee}\r)\,
{\rm e}^{i\pi}\,H_{3/2}^{(1)}(-k\eta)
= \frac{1}{\sqrt{2\,k^3}\,\ee}\,
\l(-i + k\,\eta\r)\,{\rm e}^{-i\,k\,\eta},\\
\bA_k^\prime(\eta) 
&=& -k\,\sqrt{-\frac{\pi\,\eta}{4}}\,\l(\f{\eta}{\ee}\r)\,
{\rm e}^{i\pi}\,H_{1/2}^{(1)}(-k\eta)
= -i\,\sqrt{\frac{k}{2}}\,\l(\frac{\eta}{\ee}\r)\,
{\rm e}^{-i\,k\,\eta}.
\end{eqnarray}
\end{subequations}
Then, upon using the expressions~(\ref{eq:Ji}) and~(\ref{eq:dJdphi-infl})   
as well as the above modes in the integrals~(\ref{eq:int-eta}), we 
find that the integrals [\viz $\cG_1(\bm{k_1},\bm{k_2},\bm{k_3})$ and  
$\cG_2(\bm{k_1},\bm{k_2},\bm{k_3})$] can be evaluated easily (we have
listed the results in the Appendix).
The two corresponding contributions to the three-point function can
then be calculated to be
\begin{subequations}\label{eq:Gin1}
\begin{eqnarray}
G_{\delta\phi B B(1)}(\bm{k_1},\bm{k_2},\bm{k_3}) 
&=& \frac{2\,\pi\, H_0^4}{\Mpl^2}\,
\f{(\ka+\kT)\,\l(\ka^2-\kb^2-\kc^2\right)}{\ka^3\, \kb\, \kc\,\kT^2},\\
G_{\delta\phi B B(2)}(\bm{k_1},\bm{k_2},\bm{k_3}) 
&=& -\frac{\pi\,  H_0^4}{2\,\Mpl^2}\,
\frac{1}{\ka^3\, \kb^3\, \kc^3\, \kT^2}
\l[\ka^4-2\, \ka^2 (\kb^2+\kc^2)+\kb^4+6\, \kb^2\, \kc^2+ \kc^4 \r]\nn\\
& &\times\,\biggl[\ka^3 + \kb^3\, + \kc^3 
+ 2\, \ka^2\, (\kb+\kc) + 2\, \ka\, (\kb^2+\kb \kc+\kc^2)\nn\\
& &+\,2\, \kb^2\, \kc+2\, \kb\, \kc^2\biggr],
\end{eqnarray}
\end{subequations}
where $\kT = \ka+\kb+\kc$.
The complete three-point function~$G_{\delta\phi B B}(\bm{k_1},\bm{k_2},
\bm{k_3})$ is evidently arrived at by adding the above two contributions.


\subsubsection{The case of $n=2$}

Let us now consider the case of $n=2$.
Since the scalar modes depend only on the scale factor, they remain the 
same as in the case of $n=1$ [\ie as given by Eq.~(\ref{eq:fk-ds})].
Whereas, the electromagnetic mode and its derivative are given by
\begin{subequations}
\begin{eqnarray}
\bA_k(\eta) 
&=& \sqrt{-\frac{\pi\,\eta}{4}}\,\l(\f{\eta}{\ee}\r)^2\,
{\rm e}^{3\,i\,\pi/2}\,H_{5/2}^{(1)}(-k\eta)
= \frac{-1}{\sqrt{2\,k^5}\,\ee^2}\,
\l(3 + 3\,i\,k\,\eta - k^2\,\eta^2\r)\,{\rm e}^{-i\,k\,\eta},\nn\\
\\
\bA_k^\prime(\eta) 
&=& -k\,\sqrt{-\frac{\pi\,\eta}{4}}\,\l(\f{\eta}{\ee}\right)^2\,
{\rm e}^{3\,i\,\pi/2}\,H_{3/2}^{(1)}(-k\eta)
= \frac{-\eta}{\sqrt{2\,k}\,\ee^2}\,
(1+ i\,k\,\eta)\,{\rm e}^{-i\,k\,\eta}.
\end{eqnarray}
\end{subequations}
Upon evaluating the integrals~(\ref{eq:int-eta}) using the above modes 
(in this context, see Appendix) and eventually taking the limit $\ee
\to 0$, one can obtain that
\begin{subequations}\label{eq:Gin2}
\begin{eqnarray}
G_{\delta\phi B B(1)}(\bm{k_1},\bm{k_2},\bm{k_3}) 
&=& \f{18\, \pi\, H_0^6\, a^2(\ee)}{\Mpl^2}\, 
\f{(\ka^2-\kb^2-\kc^2)}{\ka^3\, \kb^3\, \kc^3\, \kT^2}\,
\biggl[\ka^3+2\, \ka^2 (\kb+\kc)\nn\\
& &+\,2\, \ka\, (\kb^2+\kb \kc+\kc^2)
+ \kb^3+ 2\, \kb^2\, \kc+2\, \kb\, \kc^2+\kc^3\biggr],\\
G_{\delta\phi B B(2)}(\bm{k_1},\bm{k_2},\bm{k_3}) 
&=& \frac{9\, \pi\, H_0^6 a^2(\ee)}{2\,\Mpl^2}\,
\f{1}{\kb^5 \kc^5}\,\l[\ka^4-2\, \ka^2\,(\kb^2+\kc^2)
+\kb^4+6\, \kb^2\, \kc^2+\kc^4\r]\nn\\
& &\times\,\biggl\{3\, \gamma_{_{\rm E}} +3\,{\rm ln}\,(-\kT\,\ee)
-\frac{\kT^3}{\ka^3} 
- \frac{3\, \l[\ka^2-\ka\, (\kb+\kc)-\kb \kc\r]\,\kT}{\ka^3}\nn\\
& & -\,\frac{\kb\, \kc \l[3\, \ka^2\, (\kb+\kc)
+\ka\, (3\, \kb^2+8\, \kb\, \kc +3 \kc^2)
+\kb\, \kc\, (\kb+\kc)\r]}{\ka^3 \kT^2}\biggr\},\nn\\
\end{eqnarray}
\end{subequations}
where $\gamma_{_{\rm E}}$ is the Euler-Mascheroni constant.
Clearly, the complete three-point function is a sum of the above two 
contributions.

\par

A few clarifications need to made regarding the expressions we have arrived 
at above in the $n=2$ case.
To begin with, note that the two contributions in this case explicitly depend 
on~$\ee$ [cf. Eqs.~(\ref{eq:Gin2})].
The dependence arises through $a^2(\ee)$ as an overall factor and the term
${\rm ln}\,(-\kT\,\ee)$ that is encountered in the second contribution.
While the first contribution is relatively straightforward to arrive at, 
the second requires some care (when considering the $\ee\to0$ limit), 
specifically, in order to arrive at the logarithmic term. 
Later, in Sec.~\ref{sec:bnl}, we shall illustrate the amplitude
and shape of the dimensionless non-Gaussianity parameter associated with the 
three-point function $G_{\delta\phi B B}(\bm{k_1},\bm{k_2},\bm{k_3})$.
The non-Gaussianity parameter will involve the ratio of the three-point 
function and the power spectra of the magnetic field as 
well as the perturbation in the scalar field.
We shall see that the overall $a^2(\ee)$ term will be cancelled by a similar
term that arises in the power spectrum of the magnetic field.
Also, we shall find that the logarithmic term considerably enhances the 
three-point function in the flattened limit (\ie when $k_1=2\,k_2=2\,k_3$) 
leading to a characteristic shape for the non-Gaussianity 
parameter~\cite{Jain:2012vm}.


\subsection{The three-point function in the matter bounce}

Let us now turn to the evaluation of the three-point function in the 
bouncing models.
In this case, we shall assume the canonical scalar field $\phi$ to be 
a free field.
This choice is motivated by the fact that it will lead to a scale 
invariant spectrum for the perturbation in the scalar field in the 
matter bounce scenario (just as in de Sitter inflation) that we shall 
focus on~\cite{Wands:1998yp,Chowdhury:2015cma}.
Such a scalar field is governed by the following equation of motion:
\begin{equation}
\phi^{\prime\prime} + 2\,\cH\,\phi^\prime = 0.
\end{equation}
This equation can be immediately integrated to arrive at 
\begin{equation}
\phi' = \frac{C_\phi}{a^2},
\end{equation}
where $C_\phi$ is a constant of integration.
Recall that, apart from the form of the coupling function $J$, we require 
its derivative ${\rm d}J/{\rm d}\phi$  to calculate the three-point function.
Upon using the quantity $\phi'$ we have obtained above, $\d J/\d\phi$ can 
be expressed as
\begin{equation}
\frac{\d J}{\d\phi} 
= \frac{{\rm d}J}{{\rm d}\eta}\,
\frac{{\rm d}\eta}{{\rm d}\phi}
= \frac{2\,J_0\,{\bar n}\,q\,k_0^2\,a_0^{1/q}}{C_\phi}\,
\eta\,a^{{\bar n} + 2 - (1/q)}.\label{eq:dJdphi}
\end{equation}

\par

In what follows, we shall assume the background to be the matter bounce 
scenario wherein $q=1$, and we shall consider the cases ${\bar n}=1$ and
${\bar n}=3/2$. 
Since we are considering the perturbation $\delta\phi$ to 
be a massless scalar field, it essentially behaves like the tensor 
perturbation. In the case of tensor perturbations, it is well known that 
the $q=1$ case leads to scale invariant spectra~\cite{Chowdhury:2015cma}. 
It is for this reason that we shall focus on the case $q=1$ in this work.
Let us now understand the behavior of the modes.
Since the Fourier modes $f_k$ of the perturbation in the massless scalar 
field depend only on the scale factor [cf. Eq.~(\ref{eq:de-fk})], it can 
be solved for independent of the value of~${\bar n}$.
The modes $f_k$ can be arrived at by dividing the period of interest into 
two domains (over $-\infty <\eta <-\alpha\,\eta_0$ and  $-\alpha\,\eta_0<
\eta<\beta\,\eta_0$, where $\alpha =10^5$ and $\beta =10^2$) just as we had 
done in the case of the electromagnetic modes. 
In the first domain, the mode is given by the following well known matter
bounce form~\cite{Wands:1998yp,Chowdhury:2015cma}:
\begin{equation}
f_k(\eta) 
= \f{1}{\sqrt{2\,k}}\,\frac{1}{a_0\,k_0^2\,\eta^2}\,
\l(1 - \frac{i}{k\,\eta}\r)\,{\rm e}^{-i\,k\,\eta}.
\end{equation}
While in the second domain, it can be obtained to be (in this context, 
see Ref.~\cite{Chowdhury:2015cma})
\begin{equation}
f_k(\eta) 
= {\cal S}_{1k} + {\cal S}_{2k}\,g_1(k_0\,\eta),\label{eq:fk-d2}
\end{equation}
where 
\begin{subequations}
\begin{eqnarray}
{\cal S}_{1k} 
&=& \f{1}{\sqrt{2\,k}}\,
\frac{1}{a_0\,\alpha^2}\,
\l(1 + \f{i\,k_0}{\alpha\,k}\r)\,{\rm e}^{i\,\alpha\, k/k_0}
+ {\cal S}_{2k}\,g_1(\alpha),\\
{\cal S}_{2k} 
&=& \f{1}{\sqrt{2\,k}}\,\frac{1}{2\,a_0\,\alpha^2}\,\l(1+\alpha^2\r)^2\,
\l(\frac{3\,i\,k_0}{\alpha^2\,k} 
+ \frac{3}{\alpha} - \frac{i\,k}{k_0}\r)\,{\rm e}^{i\,\alpha\, k/k_0},
\end{eqnarray}
\end{subequations}
while the function $g_1(x)$ is given by
\begin{equation}
g_1(x)=\f{x}{1 + x^2} + \tan^{-1}x.\label{eq:g1}
\end{equation}
\begin{figure}[!t]
\begin{center}
\includegraphics[width=7.50cm]{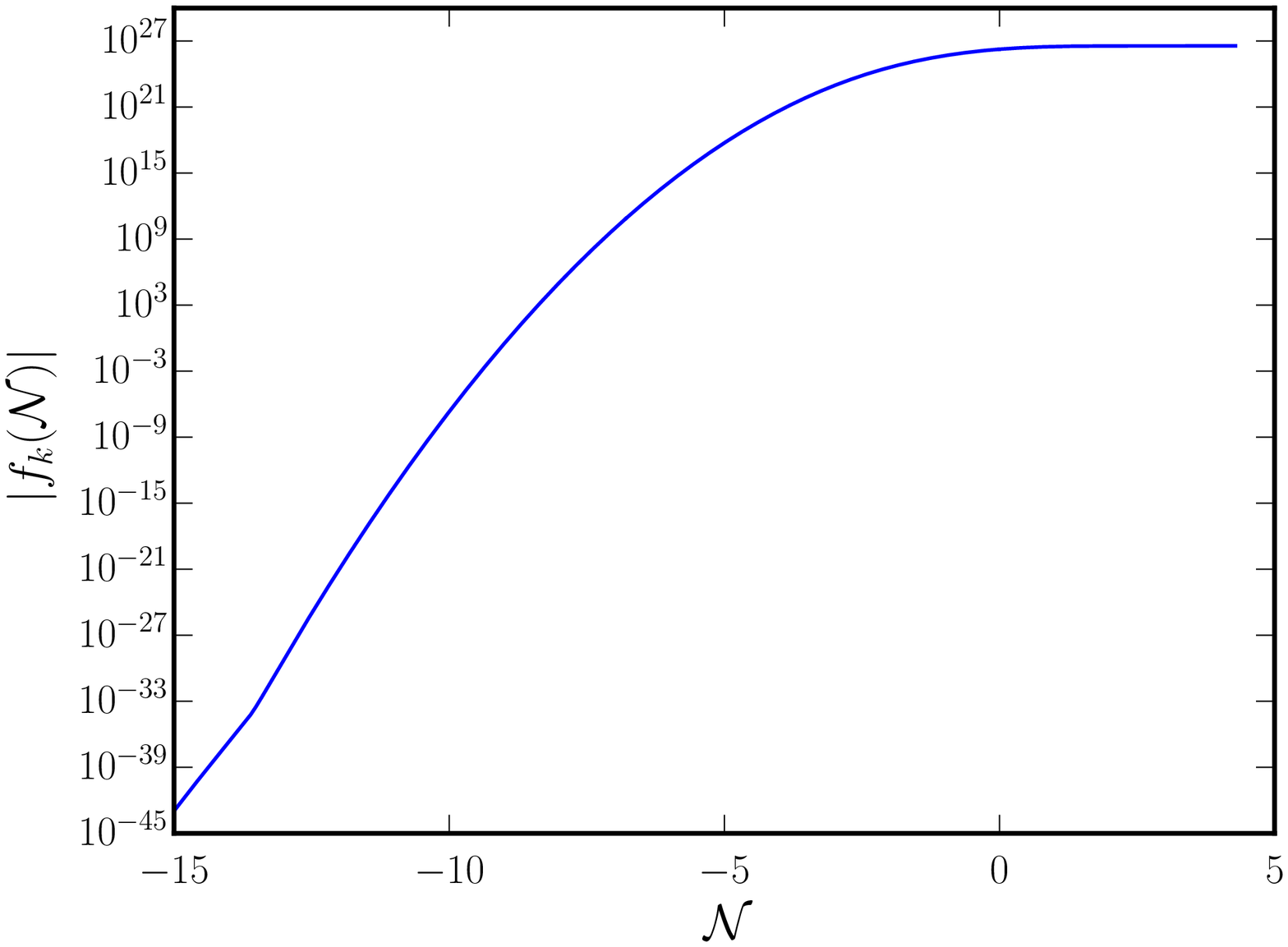}
\includegraphics[width=7.50cm]{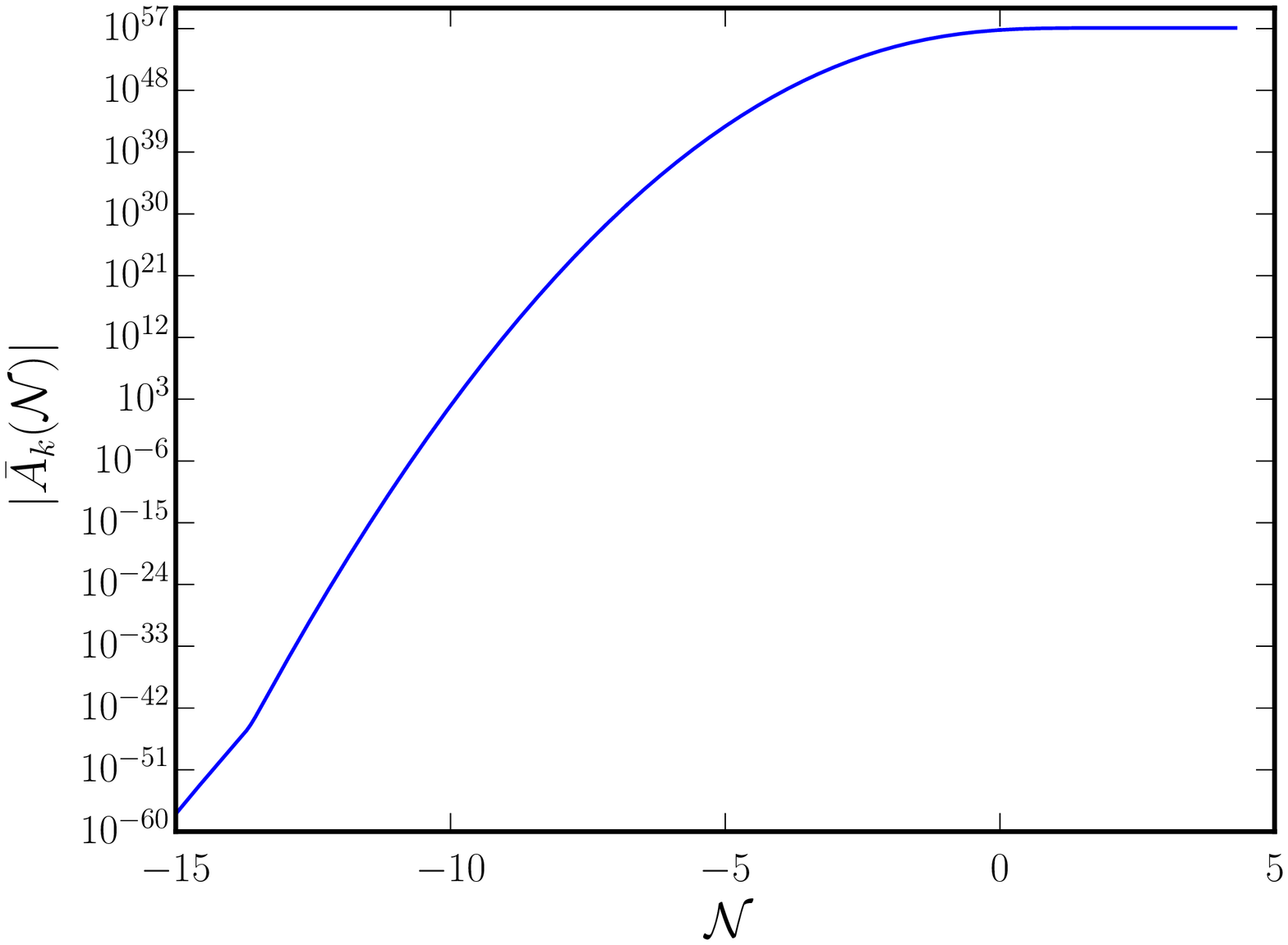}
\end{center}
\caption{
The behavior of the scalar mode $f_k$ (on the left) and the 
electromagnetic mode $\bA_k$ (on the right, for the case ${\bar n}=3/2$) in the 
matter bounce scenario (\ie when $q=1$) has been plotted as a function 
of time.
We have chosen to work with a time variable called e-N-folds that 
is convenient to describe symmetric 
bounces~\cite{Sriramkumar:2015yza}.
Note that both the modes grow as one approaches the bounce.
This property can lead to large non-Gaussianities.
These plots correspond to a typical cosmological scale with wavenumber 
$k/k_0=10^{-20}$ and we should add that modes corresponding to all the 
scales of cosmological interest behave in a similar fashion.
We should also point out that numerical analyses suggest that these 
analytical approximations match the exact numerical solutions very 
well~\cite{Sriramkumar:2015yza,Chowdhury:2015cma,
Chowdhury:2016aet}.}\label{fig:fk-Abk-b}
\end{figure}
The behavior of the mode $f_k$ is plotted in Fig.~\ref{fig:fk-Abk-b} as 
a function of so-called e-N-folds ${\cal N}$ in terms of which the scale 
factor is given by $a({\cal N})=a_0\,\exp({\cal N}^2/2)$~\cite{Sriramkumar:2015yza}, 
where $a_0$ is the value of the scale factor at the bounce.
Note that the modes $f_k$ and ${\bar A}_k$ grow strongly as one approaches
the bounce.
Such a growth may lead to large non-Gaussianities in the bouncing scenarios.


\subsubsection{The case of ${\bar n}=1$}

Let us first consider the case of ${\bar n} = 1$, as we had done in the context
of inflation.
In such a case, we have
\begin{equation}
\f{\d J}{\d\phi} 
= \f{2\,J_0\,k_0^2\,a_0^2}{C_\phi}\,\eta\,\l(1 + k_0^2\,\eta^2\r)^{2}.
\end{equation}
Since we now know the background quantities as well as the behavior 
of the modes, we have all the information required to evaluate the 
integrals characterizing the three-point function
[cf.~Eqs.~(\ref{eq:int-eta})].
In order to calculate these integrals, evidently, we can divide the period 
of interest (\ie $-\infty<\eta<\beta \,\eta_0$) into two domains ($-\infty
<\eta<-\alpha \,\eta_0$ and $-\alpha\,\eta_0<\eta<\beta \,\eta_0$) over
which we have constructed solutions for the modes. 

\par

In the first domain, for the bouncing scenario of our interest, the 
scale factor can be approximately written as $a(\eta) \simeq a_0\,
k_0^{2}\,\eta^{2}$.
Therefore, we have [cf. Eqs.~(\ref{eq:Jb}) and~(\ref{eq:dJdphi})]
\begin{equation}
J\,\frac{\d J}{\d \phi} 
\simeq \f{2\,J_0^2\,k_0^{8}\,a_0^4}{C_\phi}\,\eta^7 = J_{00}\,\eta^7,
\end{equation}
where we have set $J_{00} = 2\,J_0^2\,k_0^{8}\,a_0^4/C_\phi$.
Also, when ${\bar n}=1$, the electromagnetic modes~(\ref{eq:Ab-d1}) and their
time derivative reduce to
\begin{subequations}\label{eq:AbAbp-d1-n1}
\begin{eqnarray}
\bA_k(\eta) 
&=& \frac{-i}{\sqrt{2\,k^3}\,a_0\,J_0\,k_0^2\,\eta^3}\,
\l(1+i k\,\eta\r)\,{\rm e}^{-i\,k\,\eta},\\
\bA_k^\prime(\eta) 
&=& \frac{1}{\sqrt{2\,k^3}\,a_0\,J_0\,k_0^2\,\eta^4}\,
\l(3\,i - 3\,k\,\eta -i\,k^2\,\eta^2\r)\,{\rm e}^{-i\,k\,\eta}.
\end{eqnarray}
\end{subequations}
Now, in order to arrive at the three-point function, 
we need to evaluate the integrals~(\ref{eq:int-eta}).
After writing these equations in terms of the coupling function and the 
scalar and electromagnetic modes, we find that the integrals involved 
are mainly of the following form:
\begin{equation}
Q_m(\kT) = \int_{-\infty}^{-\alpha/k_0}\,{\rm d}\eta\,
\frac{{\rm e}^{i\,\kT\,\eta}}{\eta^m},\label{eq:Qa}
\end{equation}
which can be evaluated to be~\cite{Gradshteyn:2007}
\begin{subequations}\label{eq:Qb}
\begin{eqnarray}
Q_1(\kT) &=& i\,\pi + {\rm Ei}\l(-\frac{i\,\kT\,\alpha}{k_0}\right),\\
Q_{m+1}(\kT) &=& \frac{i\,\kT}{m}\,Q_m(\kT)
- \f{{\rm e}^{-i\,\kT\,\alpha/k_0}}{m}\,\l(-\frac{k_0}{\alpha}\r)^m,
\end{eqnarray}
\end{subequations}
with ${\rm Ei}(z)$ being the exponential integral function.
Therefore, in the first domain, the integrals~(\ref{eq:int-eta})
can be completely written in terms of the quantities~$Q_m(\kT)$
(see Appendix for the complete expressions).

\par

In the second domain, we have
\begin{equation}
J\,\frac{{\rm d}J}{{\rm d}\phi} 
= J_{01}\,\eta\l(1 + k_0^2\,\eta^2\r)^{3},
\end{equation}
where we have defined $J_{01} = 2\,J_0^2\,k_0^2\,a_0^{4}/C_\phi$.
Also, the modes and their derivatives are given by the expressions 
[cf.~Eq.~(\ref{eq:Ab-d2})]
\begin{subequations}
\begin{eqnarray}
\bA_k(\eta) 
&=& {\cal B}_{1k} + {\cal B}_{2k}\,g_1(k_0\,\eta),\\
\bA_k^{\prime}(\eta) 
&=& \f{2\,{\cal B}_{2k}\,k_0}{\l(1 + k_0^2\,\eta^2\r)^2},
\end{eqnarray}
\end{subequations}
where
\begin{subequations}
\begin{eqnarray}
{\cal B}_{1k} 
&=& \bA_k(-\alpha\,\eta_0) 
+ \f{1}{2\,k_0}\,\bA_k^{\prime}(-\alpha\,\eta_0)\,
\l(1+\alpha^2\r)^2\,g_1(\alpha),\\
{\cal B}_{2k} &=& \frac{1}{2\,k_0}\,\bA_k^{\prime}(-\alpha\,\eta_0)\,
\l(1+\alpha^2\r)^2,
\end{eqnarray}
\end{subequations}
with $\bA_k(-\alpha\,\eta_0)$ and $\bA_k'(-\alpha\,\eta_0)$ being 
obtained from the corresponding values evaluated at the end of the 
first domain [cf.~Eqs.~(\ref{eq:AbAbp-d1-n1})], while $g_1(x)$ is
given by Eq.~(\ref{eq:g1}).
On using the above expressions, we find that the 
integrals~(\ref{eq:int-eta}) in the second domain are of the following 
form:
\begin{subequations}
\begin{eqnarray}
\cG_1(\vka,\vkb,\vkc) 
&=& 4\,i\,J_{01}\,{\cal B}_{2\kb}^\ast\,{\cal B}_{2\kc}^\ast\,k_0^2\,
\int_{-\alpha/k_0}^{\beta/k_0}\,{\rm d}\eta\,
\l(\frac{\eta}{1+k_0^2\,\eta^2}\r)\,
\l[{\cal S}_{1\ka}^\ast + {\cal S}_{2\ka}^\ast\,g_1(k_0\,\eta)\r],\qquad\\
\cG_2(\vka,\vkb,\vkc) 
&=& i\,k_2\,k_3\,J_{01}\, \int_{-\alpha/k_0}^{\beta/k_0}\,{\rm d}\eta\,\eta\,
\l(1 + k_0^2\,\eta^2\r)^3\,
\l[{\cal S}_{1\ka}^\ast + {\cal S}_{2\ka}^\ast\,g_1(k_0\,\eta)\r] \nn\\
& &\times\, \l[{\cal B}_{1\kb}^\ast 
+ {\cal B}_{2\kb}^\ast\,g_1(k_0\,\eta)\r]\, 
\l[{\cal B}_{1\kc}^\ast + B_{2\kc}^\ast\,g_1(k_0\,\eta)\r].
\end{eqnarray}
\end{subequations}
These integrals can be evaluated easily in terms of elementary functions.
Thereafter, these two integrals can be combined to arrive at the total 
contribution to the three-point function from the second domain.


\subsubsection{The case of ${\bar n}=3/2$}

For the case of $q=1$ and ${\bar n} = 3/2$, which is when scale invariant 
magnetic 
fields are generated, we have
\begin{equation}
\frac{\d J}{\d\phi} 
= \frac{3\,J_0\,k_0^2\,a_0^{7/2}}{C_\phi}\,\eta\,\l(1 + k_0^2\,\eta^2\r)^{5/2}.
\end{equation}
In the first domain, for the matter bounce scenario, we therefore have 
\begin{equation}
J\,\frac{{\rm d}J}{{\rm d}\phi} \simeq 
\frac{3\,J_0^2\,k_0^{10}\,a_0^5}{C_\phi}\,\eta^9 = J_{10}\,\eta^9,
\end{equation}
where we have set $J_{10} = 3\,J_0^2\,k_0^{10}\,a_0^5/C_\phi$.
In this case, the electromagnetic modes~(\ref{eq:Ab-d1}) in the first
domain simplify to
\begin{subequations}
\begin{eqnarray}
\bA_k(\eta) 
&=& \frac{1}{J_0\,a_0^{3/2}\,k_0^3\,\sqrt{2\,k}}\,
\l(\frac{1}{\eta^3} - \frac{3\,i}{k\,\eta^4}
- \frac{3}{k^2\,\eta^5}\right)\,{\rm e}^{-i\,k\,\eta},\\
\bA_k^\prime(\eta) 
&=& \frac{1}{J_0\,a_0^{3/2}\,k_0^3\,\sqrt{2\,k}}\,
\l(\frac{15}{k^2\,\eta^6} + \frac{15\,i}{k\,\eta^5}-\frac{6}{\eta^4} 
- \frac{i\,k}{\eta^3}\right)\,{\rm e}^{-i\,k\,\eta}.
\end{eqnarray}
\end{subequations}
Using these solutions, we can compute the contribution to the 
three-point function from the first domain.
We find that the integrals~(\ref{eq:int-eta}) can be completely
expressed in terms of the functions $Q_m(\kT)$ we had introduced
in Eqs.~(\ref{eq:Qa}) and~(\ref{eq:Qb}) (see Appendix).

\par 

In the second domain, we have
\begin{equation}
J\,\frac{{\rm d}J}{{\rm d}\phi}
= J_{11}\,\eta\,\l(1 + k_0^2\,\eta^2\r)^4,
\end{equation}
where we have defined $J_{11} = 3\,J_0^2\,k_0^2\,a_0^5/C_\phi$.
Also, the electromagnetic modes are given by the following 
expressions~\cite{Chowdhury:2015cma}:
\begin{subequations}
\begin{eqnarray}
\bA_k(\eta) 
&=& {\cal C}_{1k} + {\cal C}_{2k}\,g_{3\over2}(k_0\,\eta),\\
\bA_k^{\prime}(\eta) 
&=& \frac{8\,{\cal C}_{2k}\,k_0}{\l(1 + k_0^2\,\eta^2\r)^3},
\end{eqnarray}
\end{subequations}
where the quantities ${\cal C}_{1k}$ and ${\cal C}_{2k}$ can be
written as
\begin{subequations}
\begin{eqnarray}
{\cal C}_{1k} 
&=& \bA_k(-\alpha\,\eta_0) 
+ \f{1}{8\,k_0}\,\bA_k^{\prime}(-\alpha\,\eta_0)\,\l(1+\alpha^2\r)^3\,g_{3\over2}(\alpha),\\
{\cal C}_{2k} 
&=& \frac{1}{8\,k_0}\,\bA_k^{\prime}(-\alpha\,\eta_0)\,
\l(1+\alpha^2\r)^3,
\end{eqnarray}
\end{subequations}
while the function $g_{3\over2}(x)$ is given by
\begin{equation}
g_{3\over2}(x)
=\frac{\l(5 + 3\,x^2\r)\,x}{\l(1 + x^2\r)^2} 
+ 3\,\tan^{-1}(x).
\end{equation}
Upon using the above expressions for the modes, we find that the 
integrals~(\ref{eq:int-eta}) in the second domain are of the 
following form:
\begin{subequations}
\begin{eqnarray}
\cG_1(\vka,\vkb,\vkc) 
&=& 64\,i\,J_{11}\,{\cal C}_{2\kb}^{\ast}\,{\cal C}_{2\kc}^{\ast}\,k_0^2\,
\int_{-\alpha/k_0}^{\beta/k_0}\,\f{\d\eta\,\eta}{(1 + k_0^2\,\eta^2)^2}\,
\l[{\cal S}_{1\ka}^\ast+ {\cal S}_{2\ka}^\ast\,g_1(k_0\,\eta)\r],\quad\\
\cG_2(\vka,\vkb,\vkc) 
&=& i\,k_2\,k_3\,J_{11}\, \int_{-\alpha/k_0}^{\beta/k_0}\,{\rm d}\eta\,\eta\,
\l(1 + k_0^2\,\eta^2\r)^4\,
\l[{\cal S}_{1\ka}^\ast + {\cal S}_{2\ka}^\ast\,g_1(k_0\,\eta)\r]\nn\\
& &\times\, \left[{\cal C}_{1\kb}^\ast + {\cal C}_{2\kb}^\ast\,
g_{3\over2}(k_0\,\eta)\r]\,
\left[{\cal C}_{1\kc}^\ast + {\cal C}_{2\kc}^\ast\,
g_{3\over2}(k_0\,\eta)\r].
\end{eqnarray}
\end{subequations}
These integrals can again be evaluated easily and expressed in terms 
of elementary functions and they can then be combined to arrive at
the complete three-point function. 

\par

It is useful to note here that, on comparing the amplitude of the 
contributions to the three-point function from the first and
second domains, we find that the second domain leads to a much 
larger contribution than that from the first domain in both 
the cases of ${\bar n}=1$ and ${\bar n}=3/2$.
The final expressions describing the three-point functions prove
to be quite lengthy and cumbersome.
Due to this reason, rather than write them down explicitly, we shall 
instead illustrate the behavior of the corresponding non-Gaussianity 
parameter as density plots in the next section.


\section{Amplitude and shape of the non-Gaussianity parameter}\label{sec:bnl}

Motivated by definitions of the non-Gaussianity parameters describing
the three-point auto and cross-correlations of the scalar and tensor 
perturbations~\cite{Komatsu:2010hc,Sreenath:2013xra}, we shall now define 
a non-Gaussianity parameter $\bnl$ to characterize the cross-correlation 
between the magnetic field and the perturbation in the scalar field.
As we shall see, the parameter will prove to be a dimensionless quantity 
involving the ratio of the cross-correlation and the power spectra of the 
scalar perturbation and the magnetic field.
The parameter captures the amplitude and shape of the three-point function,
and we should clarify that it has a form very similar to the parameters 
defined earlier in this context~\cite{Caldwell:2011ra,Jain:2012ga,Jain:2012vm}.
In this section, we shall arrive at the expression for the non-Gaussianity 
parameter by following the same procedure as was used to arrive at similar 
parameters for the three-point functions involving the scalar and tensor 
perturbations~\cite{Sreenath:2013xra}.
With the definition at hand, we shall evaluate the parameter for the 
inflationary and bouncing scenarios of our interest.

\par

The amplitude of the non-Gaussianity in the local model 
for the scalar three-point function is usually parameterized in terms 
of a parameter $\fnl$ which, in this model, coincides with the bispectrum 
scaled by products of the power spectra.
Here, we generalize that analogously to a non-Gaussianity parameter 
$\bnl$ which we define through the following relation:
\begin{eqnarray}
\hat{B}_{i\,\bm{q}}(\eta) 
= \hat{B}_{i\,\bm{q}}^{({\rm G})}(\eta) 
+ \frac{\bnl}{2\,\Mp}\, 
\int \f{\d^3\bm{p}}{(2\,\pi)^{3/2}}\,
\hat{\delta\phi}_{\bm{q}-\bm{p}}(\eta)\,
\hat{B}_{i\,{\bm p}}^{({\rm G})}(\eta),
\end{eqnarray}
where $\hat{B}_{i\,\bm{q}}$ is the Fourier mode of the actual magnetic 
field and $\hat{B}_{i\,\bm{q}}^{({\rm G})}$ indicates the Fourier mode of
its Gaussian part, while, as usual, $\hat{\delta\phi}_{\bm{q}-\bm{p}}$ 
refers to Fourier mode of the perturbation in the scalar field which 
has already been assumed to be Gaussian.
We can evaluate the three-point function of our interest, \viz
$\langle \hat{\delta\phi}_{\bm{k_1}}\,\hat{B}^i_{\bm{k_2}}\,
\hat{B}_{i\,\bm{k_3}}\rangle$, upon using the above definition 
of $\hat{B}_{i\,{\bm q}}$ and Wick’s theorem which applies
to Gaussian operators. 
On making use of the definition~(\ref{eq:cc}) of 
$G_{\delta\phi B B}(\bm{k_1},\bm{k_2},\bm{k_3})$ and inverting
the resulting expression, we obtain the following expression
for $\bnl$:
\begin{eqnarray}\label{eq:bnl}
\bnl\l(\bm{k_1},\bm{k_2},\bm{k_3}\r) 
&=& \f{1}{16\,\pi^5}\, \f{J^2(\ee)}{a^2({\ee})}\,
\biggl[\ka^3\,\kb^3\,\kc^3\,
G_{\delta\phi B B}(\bm{k_1},\bm{k_2},\bm{k_3})\biggr]\nn\\
& &\times\, \biggl\{\cP_{\delta\phi}(\ka)\left[\kc^3\,\psb(\kb)
+ \kb^3\,\psb(\kc)\right]\biggr\}^{-1}.
\end{eqnarray}
In this expression, $\psb(k)$ is the power spectrum of the magnetic 
field we have discussed earlier, while $\cP_{\delta\phi}(k)$ is the 
power spectrum of the scalar field, given by
\begin{equation}
\cP_{\delta\phi}(k) 
= \f{k^3}{2\,\pi^2\,\Mp^2}\,\vert\,f_{k}\vert^2,
\end{equation}
with the right hand side to be evaluated as $\ee\to 0$ in the context
of inflation and at $\eta=\beta\,\eta_0$ in the context of the
bouncing models. 
Due to their dual nature~\cite{Wands:1998yp}, both de 
Sitter inflation and matter bounce are expected to lead to scale invariant 
spectra for the perturbation in the scalar field.
In de Sitter inflation, the spectrum is given by the well known scale
invariant form
\begin{equation}
\cP_{\delta\phi}(k)=\f{H_0^2}{4\,\pi^2\,\Mp^2}.\label{eq:pssi}
\end{equation}
In the matter bounce, using the modes~(\ref{eq:fk-d2}), it can be 
determined to be~\cite{Chowdhury:2015cma}
\begin{equation}
\cP_{\delta\phi}(k)=\f{9\, k_0^2}{16\,a_0^2\,\Mpl^2}.
\label{eq:pssb}
\end{equation}

\par

Let us first consider the amplitude and shape of the non-Gaussianity 
parameter $\bnl$ for an arbitrary triangular configuration of the 
wavevectors $\vka$, $\vkb$ and $\vkc$ in the context of inflation.
In de Sitter inflation, for the case of $n=1$, from the two 
contributions~(\ref{eq:Gin1}) to the three-point function and the power 
spectra~(\ref{eq:psbi}) and~(\ref{eq:pssi}),
the parameter $\bnl$ can be obtained to be
\begin{eqnarray}
\bnl(\bm{k_1},\bm{k_2},\bm{k_3}) 
&=& -\frac{1}{2\, \kb^2\, \kc^2\, (\kb+\kc)\, \kT^2}\,
\biggl\{4\, \kb^2\, \kc^2\, (\ka+\kT)\, \l(-\ka^2+\kb^2+\kc^2\r) \nn\\
& &+\, \biggl[\ka^4-2\, \ka^2\, \l(\kb^2+\kc^2\r)
+\kb^4+6\, \kb^2 \kc^2+\kc^4\biggr]\,
\biggl[\ka^3+2\, \ka^2\, (\kb+\kc)\nn\\
& &+\,2\, \ka\, \l(\kb^2+ \kb \kc + \kc^2\r)
+ \kb^3+2\, \kb^2\, \kc+2\, \kb\, \kc^2+\kc^3\biggr]\biggr\}.
\end{eqnarray}
Note that, in the squeezed limit, wherein $\ka \to 0$ and $\kb=\kc$, 
we have $\bnl = -4$.
For the case of $n=2$, the non-Gaussianity parameter can be obtained 
[from Eqs.~(\ref{eq:Gin2}), (\ref{eq:psbi}) and (\ref{eq:pssi})] to be
\begin{eqnarray}
\bnl(\bm{k_1},\bm{k_2},\bm{k_3}) 
&=& \frac{1}{2\, \kb^2\, \kc^2\, \l(\kb^3+\kc^3\r)\, \kT^2}\,
\biggl\{4\,\kb^2\, \kc^2\, \l(\ka^2-\kb^2-\kc^2\r)\, 
\biggl[\ka^3+2\, \ka^2\, (\kb+\kc)\nn\\
& &+\, 2\, \ka\, \l(\kb^2+\kb\, \kc+\kc^2\r)+\kb^3+2\, \kb^2\, \kc
+2\, \kb\, \kc^2+\kc^3\biggr]\,
-\biggl[\ka^4+\kb^4+\kc^4\nn\\
& &-\,2\, \ka^2 \l(\kb^2+\kc^2\r) + 6\, \kb^2\, \kc^2\biggr]\, 
\biggl[-3\, \gamma_{_{\rm E}}\,  \ka^3\, \kT^2 
- 3\,\ka^3\,\kT^2\, {\rm ln}\,(-\kT\,\ee)\nn\\
& &+\,\kb\, \kc\, \biggl(3\, \ka^2\, (\kb+\kc) 
+ \ka\, \l(3\, \kb^2+8\, \kb\, \kc+3\, \kc^2\r) 
+ \kb\, \kc\, (\kb+\kc)\biggr)\nn\\
& &+\,3\, \biggl(\ka^2-\ka (\kb+\kc) 
-\kb\, \kc\biggr)\, \kT^3 + \kT^5\biggr]\biggr\}.
\end{eqnarray}
In the squeezed limit, we have $\bnl = -8$.
In the next section, we shall discuss the properties of $\bnl$ in 
the squeezed limit in more detail.

\par 

For the matter bounce scenario and the two cases of ${\bar n}=1$ and 
${\bar n}=3/2$
that we had considered, we can arrive at the non-Gaussianity parameter 
$\bnl$ from the various expressions we had obtained earlier and the 
corresponding power spectra.
However, as the resulting expressions are too lengthy and cumbersome, 
we do not explicitly write them down here.
We shall plot them below and compare them with the results in inflation.

\par

\begin{figure}[!t]
\begin{center}
\includegraphics[width=7.50cm]{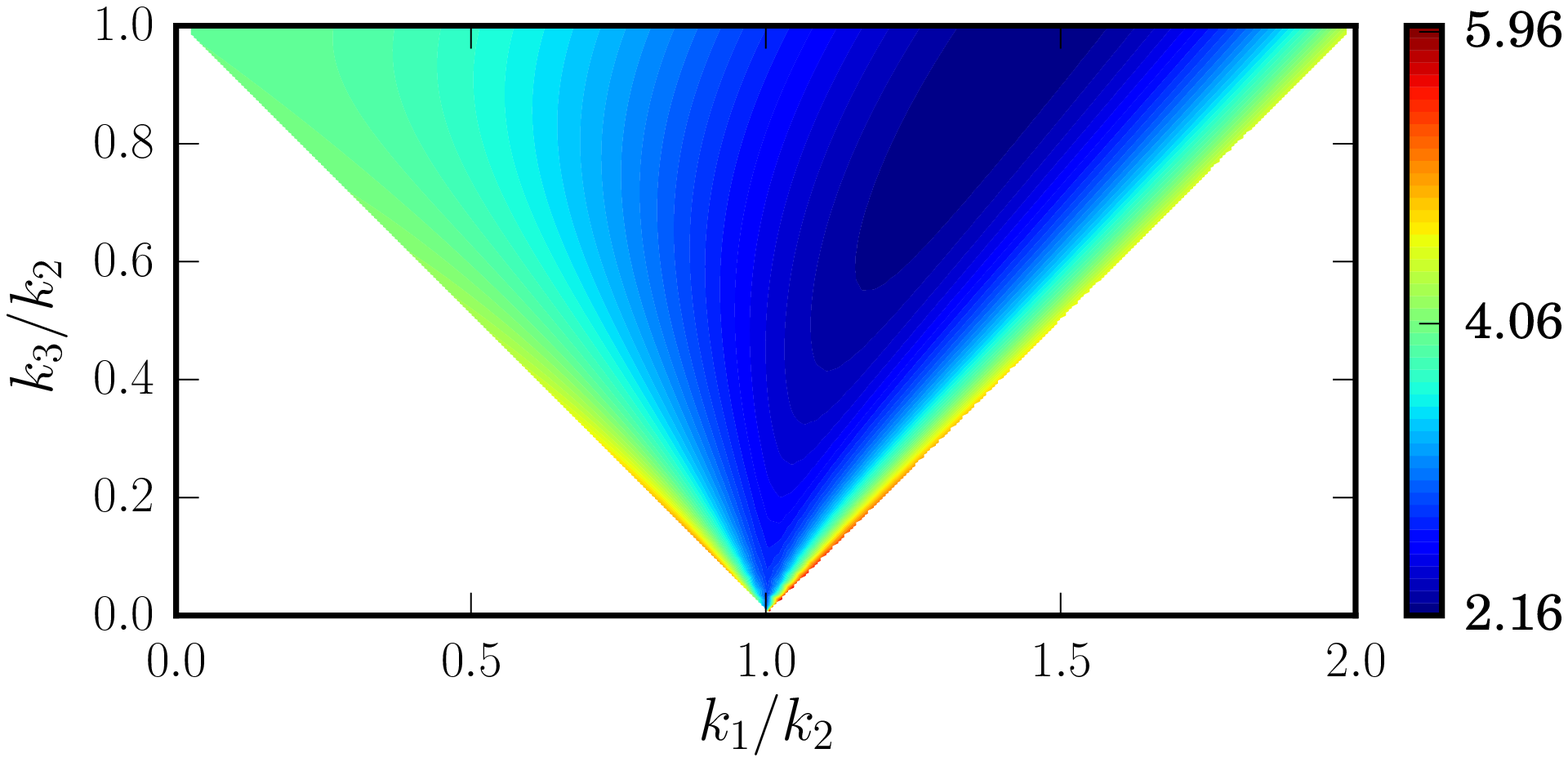}
\includegraphics[width=7.50cm]{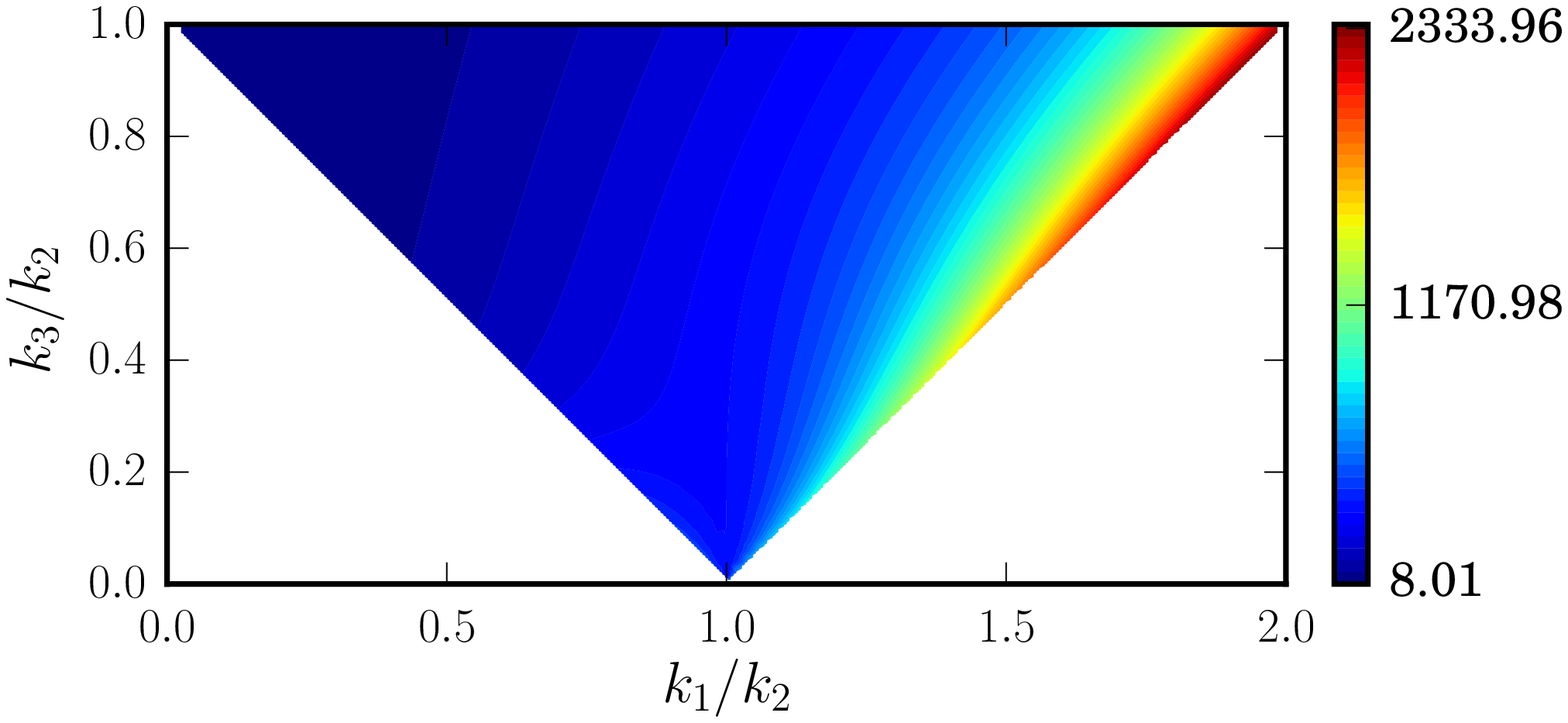}
\end{center}
\caption{The dimensionless non-Gaussianity parameter~$\bnl$ 
arising in de Sitter inflation has been illustrated as a 
density plot for an arbitrary triangular configuration of 
wavenumbers for the two cases of~$n$ that we had considered 
(with $n=1$ on the left and $n=2$ on the right).
Recall that the $n=2$ case leads to a scale invariant spectrum 
for the magnetic field.
The amplitude as well as the shape of the parameter is considerably
different in the two cases.
Moreover, in the case $n=2$, the parameter is considerably enhanced
in the flattened limit (due to the logarithmic term depending on $\ee$
that arises in the three-point function), \ie when
$k_1=2\,k_2=2\,k_3$~\cite{Jain:2012ga,Jain:2012vm}.
Further, $\bnl$ tends to $-4$ and $-8$ when $n=1$ and $n=2$ in the 
squeezed limit (\ie as $k_1\to 0$), respectively, indicating that 
the cross-correlation satisfies the consistency relation, a point 
which we shall discuss in some generality in the next section.
We have chosen $\ee$ such that the pivot scale of
$k_\ast=0.002\,{\rm Mpc}^{-1}$ leaves the Hubble radius $50$ e-folds before
the end of inflation.
In this case, note that, we can write 
${\rm ln}\,(-\kT\,\ee)={\rm ln}\,(\kT/k_\ast)-50$.}\label{fig:bnl-i}
\end{figure}
\begin{figure}[!t]
\begin{center}
\includegraphics[width=7.50cm]{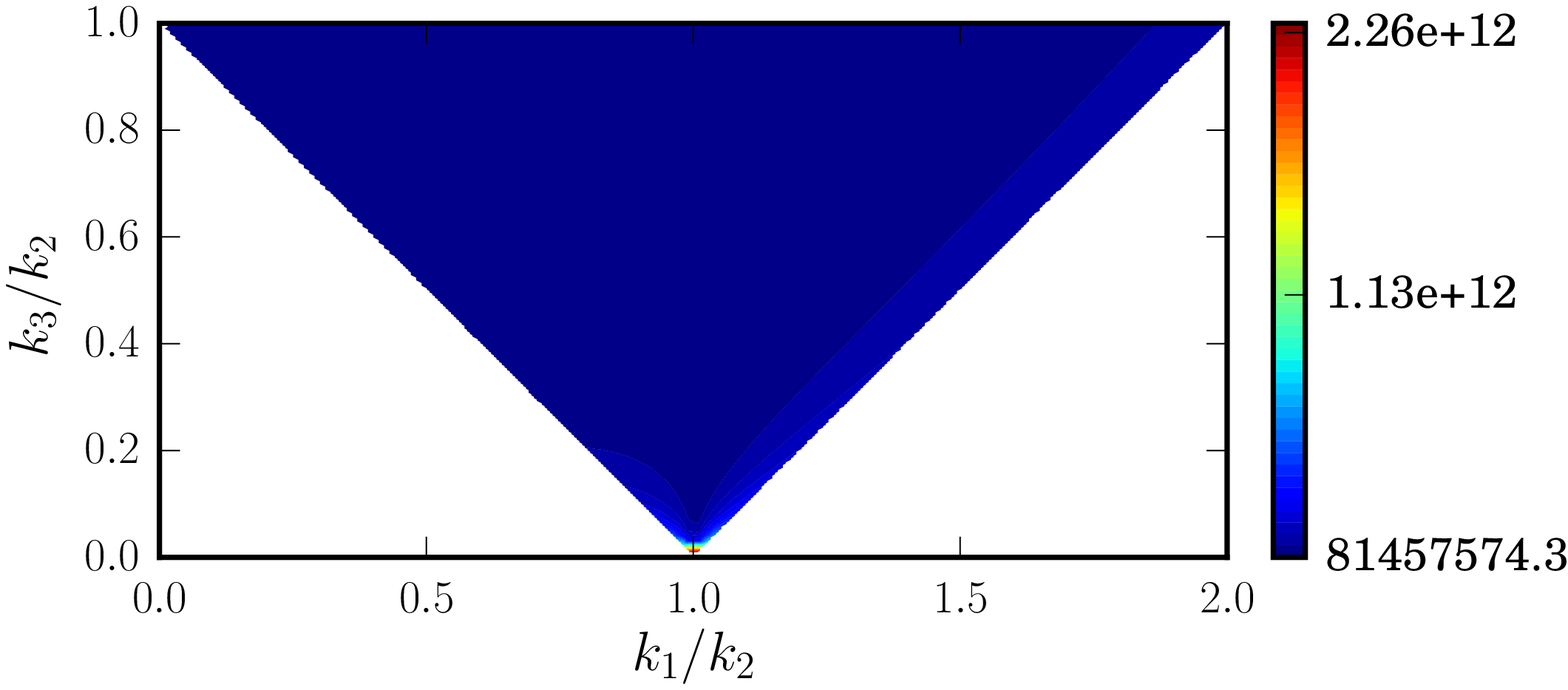}
\includegraphics[width=7.50cm]{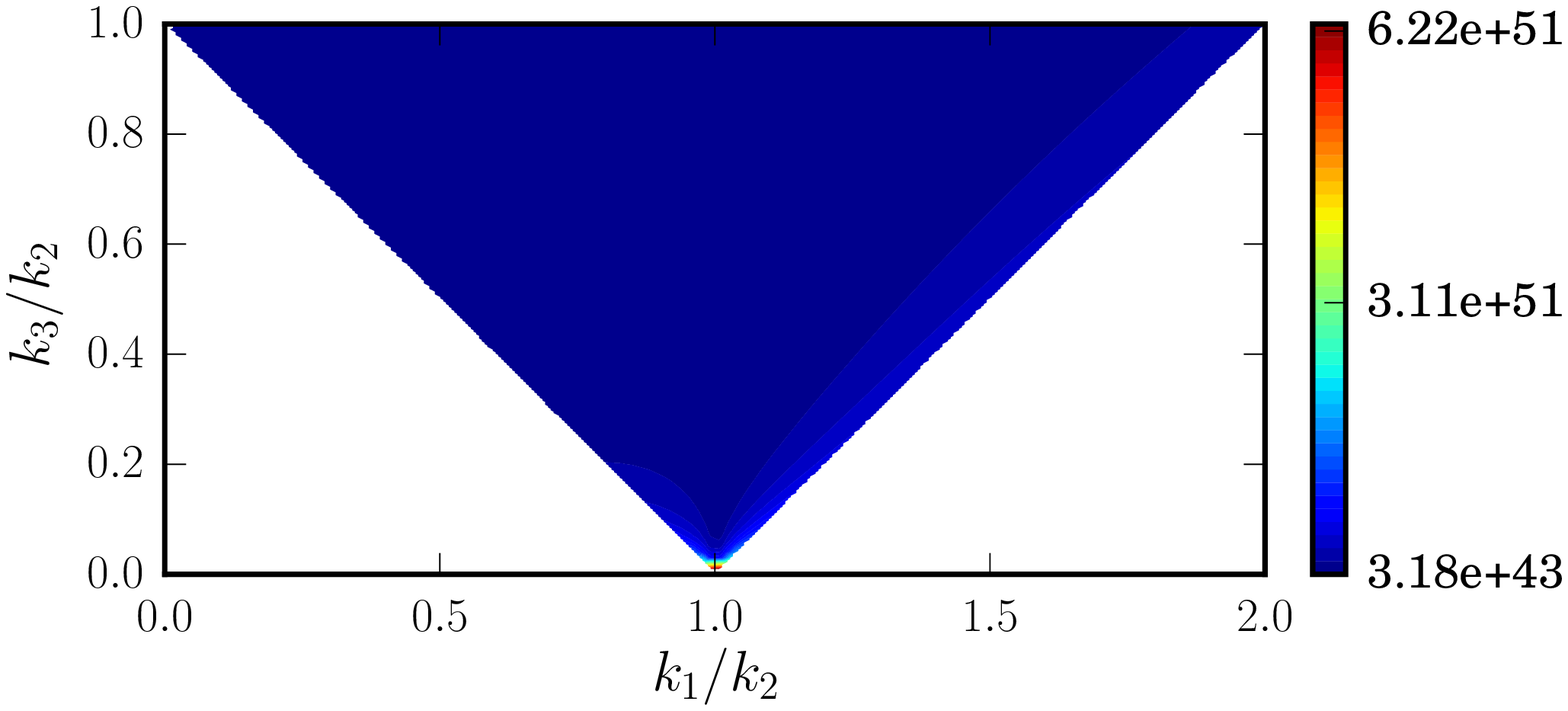}
\end{center}
\caption{The dimensionless non-Gaussianity parameter $\bnl$ arising 
in the matter bounce scenario has been illustrated in the form of 
density plots as in the previous figure for the two cases of ${\bar n}$ we 
had discussed (${\bar n}=1$ on the left and ${\bar n}=3/2$ on the right).
When compared with the results in the inflationary case, two points 
are evident.
Firstly, the shape of the parameter proves to be considerably more 
homogeneous.
This is primarily because of the fact that all the scales 
of cosmological interest are much smaller than the bounce 
scale~$k_0$, and hence the simple bouncing scenario we are 
considering here does not strongly discriminate between 
the wavenumbers of interest.
Secondly, the amplitude of $\bnl$ is considerably larger.
This can be completely attributed to the forms of the coupling 
functions that we have considered.
Moreover, in the ${\bar n}=3/2$ case, there arises a strong red tilt with,
say, $\bnl$ behaving as $k^{-2}$ in the equilateral limit, which 
leads to even larger values (compared to the ${\bar n}=1$ case, 
where it is nearly scale invariant) over cosmological scales.
In contrast, in de Sitter inflation, $\bnl$ is nearly scale invariant 
in all the limits for both the values of $n$ that we have considered.}
\label{fig:bnl-b}
\end{figure}
In Figs.~\ref{fig:bnl-i} and~\ref{fig:bnl-b}, we have illustrated 
the non-Gaussianity parameter $\bnl$ as a density plot for an 
arbitrary triangular configuration of the wavevectors $\vka$, 
$\vkb$, and $\vkc$, as is often done for the scalar non-Gaussianity 
parameter $\fnl$ (see, for instance, Ref.~\cite{Komatsu:2010hc}).
In these plots, the $x$-axis corresponds to the ratio of the amplitudes 
of $\vka$ and $\vkb$, while the $y$-axis corresponds to the ratio of the 
amplitudes of $\vkc$ and $\vkb$.
For the case of three-point functions involving perturbations of similar 
nature, such as the scalar and tensor bispectra, due to the symmetrical 
nature of these three-point functions, it is sufficient to construct 
these density plots over the domain $0.0 <\ka/\kb< 1.0$ and $0.5<\kc/\kb
<1.0$.
However, for the cases of three-point functions involving one mode of 
a nature dissimilar from the other two, such as in the case of the 
three-point function with one scalar mode and two electromagnetic modes 
that we are studying here, we find that the above ranges of ratios of 
the wavenumbers do not adequately represent all the various combinations 
of wavenumbers that can arise in such a context.
Therefore, in Figs.~\ref{fig:bnl-i} and~\ref{fig:bnl-b}, we have plotted 
the $\bnl$ parameter over the ranges $0.0<\ka/\kb<2.0$ and $0.0<\kc/\kb
<1.0$.
These values of the ratios of wavenumbers can satisfactorily represent 
the entire range of triangular configuration of wavevectors of our
interest.

\par

A few points need to be stressed regarding the results that have been 
illustrated in Figs.~\ref{fig:bnl-i} and~\ref{fig:bnl-b}.
Let us first discuss the case of de Sitter inflation.
To begin with, note that, in this case, the amplitude as well as the shape of the 
non-Gaussianity parameter $\bnl$ is considerably different depending on
whether $n=1$ or $n=2$.
Also, the amplitude in the $n=2$ case is considerably larger due to
the appearance of the $\ln\, (-\kT\,\ee)$ term, with the maximum values
of the parameter arising in the flattened limit wherein
$\ka=2\,\kb=2\,\kc$.
Moreover, as we already mentioned, in the squeezed limit wherein $\ka\to0$
and $\kb=\kc$, we have $\bnl=-4$ when $n=1$ and $\bnl=-8$ when $n=2$.
This is essentially the so-called consistency relation which we shall 
establish in the next section for an arbitrary $n$ in the case of inflation.
In contrast, the non-Gaussianity parameter seems to have primarily the same 
shape for ${\bar n}=1$ and ${\bar n}=3/2$ in the matter bounce case.
This can be attributed to the fact that, as the wavenumbers of cosmological
interest are much smaller than the bounce scale, the matter bounce 
scenario is unable to strongly discriminate between these modes.
However, the amplitude of $\bnl$ in the matter bounce is considerably larger
than in de Sitter inflation, which is possibly because of the form of the
coupling function that we have considered.
While, for ${\bar n}=1$, $\bnl$ is nearly scale invariant, say, in the equilateral 
limit, we find that the parameter has a strong red tilt when ${\bar n}=3/2$ (with 
$\bnl$ behaving as $k^{-2}$), leading to rather large values for the small 
wavenumbers (say, for $10^{-20}< k/k_0 <10^{-15}$) corresponding to 
cosmological scales.
We shall discuss the implications of this result in the concluding section.


\section{The three-point function in the squeezed limit and the 
consistency relation}\label{sec:sq-cons}

One of the most important characteristics of three-point functions 
is their behavior in the squeezed limit of the three wavenumbers, 
\viz when one of the three wavenumbers is much smaller than the 
other two, or, equivalently, when one of the three wavelengths is
much longer than the other two.
For the case of inflation, in this limit, the longest wavelength 
mode freezes on super Hubble scales and therefore acts as a background 
for the smaller wavelength modes.
Consequently, the three-point function can be expressed in terms of 
the two-point functions involving the perturbations through the 
so-called consistency relation~\cite{Maldacena:2002vr,Creminelli:2004yq}.
In this section, we shall obtain the consistency relation for the 
three-point function involving the magnetic field and the scalar 
perturbation in the case of de Sitter inflation.
We shall also discuss its validity in the bouncing scenario.

\par

Let us first consider the case of de Sitter inflation.
In the squeezed limit, \ie when $\ka \to 0$ and, say, $\vkb = -\vkc = \vk$,  
the scalar mode with wavenumber $\ka$ can be considered to have exited 
the Hubble radius and hence its amplitude can be treated as a constant.
Therefore, the mode $f_k$ can be extracted out of the 
integrals~(\ref{eq:int-eta}).
For the coupling function $J(\phi)$ that we had considered, given the
modes~(\ref{eq:Abki}), we obtain that, for arbitrary $n$~\cite{Gradshteyn:2007},
\begin{subequations}\label{eq:cG-i-sq}
\begin{eqnarray}
\cG_1(\vka,\vk,-\vk) 
&=& \frac{i}{\Mp}\,\int_{-\infty}^{\ee}\,{\rm d}\eta\,
J^2(\eta)\,f_{\ka}^\ast(\eta)\,{\bar A}_{k}^{\prime\ast 2}(\eta)\nn\\
&\simeq& \frac{i}{\Mp}\,f_{\ka}^\ast(\ee)\,
\int_{-\infty}^{\ee}\,{\rm d}\eta\, J^2(\eta)\,
{\bar A}_{k}^{\prime\ast 2}(\eta)\nn\\
&=& \frac{i\,\pi\,k^2\,\ee^2\,{\rm e}^{-i\,n\,\pi}}{8\,\Mp}\,
f_{\ka}^\ast(\ee)\nn\\
& &\times\,\biggl\{\l[H_{n-1/2}^{(2)}(-k\,\ee)\r]^2
- H_{n-3/2}^{(2)}(-k\,\ee)\,H_{n+1/2}^{(2)}(-k\,\ee)\biggr\},
\label{eq:cG-1-infl-sq}\\
\cG_2(\vka,\vk,-\vk) 
&=& \frac{i}{\Mp}\,k^2\,\int_{-\infty}^{\ee}\,{\rm d}\eta\,
J^2(\eta)\,f_{\ka}^\ast(\eta)\,{\bar A}_{k}^{\ast 2}(\eta)\nn\\
&\simeq& \frac{i\,k^2}{\Mp}\,f_{\ka}^\ast(\ee)\,\int_{-\infty}^{\ee}\,{\rm d}\eta\,
J^2(\eta)\,{\bar A}_{k}^{\ast 2}(\eta) \nn\\
&=& \frac{i\,\pi\,k^2\,\ee^2\,{\rm e}^{-i\,n\,\pi}}{8\,\Mp}\,f_{\ka}^\ast(\ee)\nn\\
& &\times\,\biggl\{\l[H_{n+1/2}^{(2)}(-k\,\ee)\r]^2
- H_{n-1/2}^{(2)}(-k\,\ee)\,H_{n+3/2}^{(2)}(-k\,\ee)\biggr\}.
\label{eq:cG-2-infl-sq}
\end{eqnarray}
\end{subequations}
Adding the contributions due to these two integrals, the three-point 
function in the squeezed limit can be written as
\begin{eqnarray}
G_{\delta\phi BB}(\vka,\vk,-\vk) 
&=& \frac{-i\,\pi^3\,H_0^2\,\ee^5\,k^4}{2\,\Mp^2}\,
\left|f_{\ka}^\ast(\ee)\right|^2\nn\\
& &\times\,\Biggl(\left[H_{n+1/2}^{(1)}(-k\,\ee)\right]^2\,
\biggl\{\left[H_{n-1/2}^{(2)}(-k\,\ee)\right]^2
+ \left[H_{n+1/2}^{(2)}(-k\,\ee)\right]^2\nn\\
& &-\, H_{n-3/2}^{(2)}(-k\,\ee)\,H_{n+1/2}^{(2)}(-k\ee) 
- H_{n-1/2}^{(2)}(-k\,\ee)\,H_{n+3/2}^{(2)}(-k\,\ee)\biggr\}\nn\\
& &- \l[H_{n+1/2}^{(2)}(-k\,\ee)\r]^2\, 
\biggl\{\left[H_{n-1/2}^{(1)}(-k\,\ee)\right]^2 
+ \left[H_{n+1/2}^{(1)}(-k\,\ee)\right]^2\nn\\
& &-\, H_{n-3/2}^{(1)}(-k\,\ee)\,H_{n+1/2}^{(1)}(-k\,\ee) 
- H_{n-1/2}^{(1)}(-k\,\ee)\,H_{n+3/2}^{(1)}(-k\,\ee)\biggr\}\Biggr).\nn\\
\end{eqnarray}
Using this result, the expressions for the power spectra of the scalar 
field and the magnetic field as well as the behavior of Hankel functions 
$H_\nu(x)$ for small values of $x$, we can obtain the non-Gaussianity 
parameter in the squeezed limit to be~\cite{Gradshteyn:2007}
\begin{equation}
\lim_{\ka\to0}\bnl(\vka,\vk,-\vk) = 2\,n_{_{\rm B}} - 8,\label{eq:cr}
\end{equation}
where $n_{_{\rm B}} = (4 - 2\,n)$ is the spectral index of the power
spectrum of the magnetic field.
This implies that, in the squeezed limit, $\bnl=-4$ for $n=1$ and 
$\bnl=-8$ for $n=2$, results that we had already arrived at in the 
previous section.

\par

In the bouncing scenario under consideration, we find that the scalar 
mode strongly grows as one approaches the bounce and, in fact, also
exhibits a slow growth even after the bounce (in this context, see
Fig.~\ref{fig:fk-Abk-b}).
Specifically, in contrast to its behavior in de Sitter inflation, the 
scalar mode $f_k$ does not become a constant at late times.
Therefore, it can be expected that, contrary to the reduced 
integrals~(\ref{eq:cG-i-sq}) that we had obtained in the 
squeezed limit in the inflationary context, the corresponding
integrals in bounces would involve all the three modes.
Consequently, the consistency relation~(\ref{eq:cr}) that is valid in 
the case of de Sitter inflation may not hold true in the bouncing model.
On evaluating the three-point function and the non-Gaussianity parameter 
in the bounce, we find that the consistency relation is indeed violated.
The violation of the consistency relation corresponding to the tensor
bispectrum in a matter bounce was observed in a previous 
work~\cite{Chowdhury:2015cma}, and it is interesting to note that a 
similar behavior is exhibited by the three-point function involving one 
scalar and two electromagnetic modes as well.
This difference in the behavior of the non-Gaussianity parameter between 
de Sitter inflation and matter bounce scenarios, despite the similarities 
in the two-point functions obtained in these two models, can potentially 
serve as a discriminator between the inflationary paradigm and the 
bouncing scenarios.


\section{Discussion}\label{sec:dis}

The magnetic fields existing on various scales in the universe are 
considered to have originated from a primordial seed field. 
Although the generation of such fields via the inflationary mechanism 
has been well studied, there have been far fewer endeavors to investigate 
the origin of the primordial magnetic fields in bouncing models. 
One of the most important signatures of such fields would be the extent 
of non-Gaussianity associated with the cross-correlations between these 
fields and the scalar perturbations. 
These three-point functions have been evaluated in certain inflationary 
scenarios and it has been found that the corresponding non-Gaussianity 
parameter can be expected to be quite large. 

\par

In the matter bounce scenario of our interest, the modes grow as one 
approaches the bounce, and grow very slowly thereafter. 
Due to the enhancement in the amplitude of the modes, it can be expected 
that the non-Gaussianities associated with the three-point functions 
would be very large.
In a previous work~\cite{Chowdhury:2015cma}, it was shown that irrespective 
of the growth of the tensor modes in a matter bounce, the corresponding 
tensor non-Gaussianities are rather small and the consistency relation for 
the tensor bispectrum is violated.
In contrast, in this work, we find that the non-Gaussianity parameter 
associated with the cross-correlations involving the primordial magnetic 
fields and the scalar perturbations is very large and, in fact, is much 
larger than what is expected in the de Sitter inflationary scenario.
Further, the corresponding consistency relation is violated in the bouncing 
model.
It should be noted that the bouncing scenario that we have considered 
here suffers from considerable backreaction in the vicinity of the 
bounce~\cite{Sriramkumar:2015yza,Chowdhury:2016aet}, which can possibly 
be responsible for the large non-Gaussianities that we encounter.
This issue of circumventing the backreaction in the proximity of the 
bounce is non-trivial and needs to be investigated in more detail.

\par 

Recall that we have restricted ourselves to the matter bounce scenario.
Evidently, other combinations of the parameters ${\bar n}$ and $q$ can also lead 
to scale invariant magnetic power spectra of relevant amplitude.
It would be interesting to examine whether other scenarios that result in
similar amplitude for the scale invariant power spectrum also lead to similar
results for the non-Gaussianities.
Note that the extent of non-Gaussianities depends on $J(\phi)$ as well as
$\d J/\d \phi$, with the latter depending on $\phi'$ [cf. Eq.~(\ref{eq:dJdphi})].
Therefore, it is important to explore a wide variety of coupling functions
$J(\phi)$, some of which may lead to considerably different levels of 
non-Gaussianities.
It seems difficult to be able to make generic remarks regarding the
results one can obtain from different forms of couplings.
Therefore, one may have to examine the effects of each type of coupling separately.
We believe that it may be challenging to simultaneously produce magnetic 
fields of observable strengths in bounces while also ensuring that the levels of
non-Gaussianities remain as small as encountered in inflation. 
We are curently systematically exploring these issues.


\section*{Acknowledgements}

This work was initiated during visits by DC and LS to the Department 
of Physics and Astronomy, Johns Hopkins University, Baltimore, Maryland, 
U.S.A. under the aegis of The Indo-US Science and Technology Forum 
grant~IUSSTF/JC-Fundamental Tests of Cosmology/2-2014/2015-16.
DC would like to thank the Indian Institute of Technology Madras, Chennai, 
India, for financial support through half-time research assistantship.
LS also wishes to thank the Indian Institute of Technology Madras, 
Chennai, India, for support through the Exploratory Research Project
PHY/17-18/874/RFER/LSRI.
This work was supported at Johns Hopkins University by 
NSF Grant No. 1519353, NASA NNX17AK38G, and the Simons Foundation.
We wish to thank Rajeev Jain for comments on the manuscript.


\appendix

\section{Appendix:~Evaluation of integrals}


In this appendix, we shall provide the results for the integrals 
$\cG_1(\bm{k_1},\bm{k_2},\bm{k_3})$ and  $\cG_2(\bm{k_1},\bm{k_2},
\bm{k_3})$ [cf.~Eqs.~(\ref{eq:int-eta})] evaluated in de Sitter inflation 
and in the first domain in the matter bounce scenario of our interest.

\par

Let us first consider the $n=1$ case in de Sitter inflation.
In this case, we find that the integrals can be evaluated to be
\begin{subequations}
\begin{eqnarray}
\cG_1 (\vka,\vkb,\vkc) 
&=& \frac{i\,H_0\,\sqrt{\kb\,\kc}
{\rm e}^{i\,\kT\,\ee}}{\sqrt{8\,\ka^3}\,\Mp\,\kT^2}\,
\l(-i\,\ka\,\kT\,\ee+2\,\ka+\kb+\kc\r),\\
\cG_2 (\vka,\vkb,\vkc) 
&=& \frac{H_0\,k_2\,k_3\,{\rm e}^{i\,\kT\,\ee}}
{\sqrt{8\,\ka^3\,\kb^3\,\kc^3}\,\Mp}\,
\biggl\{\frac{1}{\ee}-\frac{\ka\,\kb\,\kc\,\ee}{\kT} \nn\\
& &-\,\frac{i}{\kT^2}\,
\l[\ka^2\,(\kb+\kc)+\ka\,\l(\kb^2+4\,\kb\,\kc+\kc^2\r)
+\kb\,\kc\,(\kb+\kc)\r]\biggr\}.\nn\\
\end{eqnarray}
\end{subequations}
In the case of $n=2$, we find that the integrals are given by
\begin{subequations}
\begin{eqnarray}
\cG_1 (\vka,\vkb,\vkc) 
&=& \frac{i\,H_0\,{\rm e}^{i\,\kT\,\ee}}{\sqrt{8\,\ka^3\,\kb\,\kc}\,\Mp}
\biggl[\frac{i}{\ee}-\frac{i\,\ka\,\kb\,\kc\,\ee}{\kT} \nn\\
&+& \frac{\ka^2\,(\kb+\kc)+\ka\,\left(\kb^2+4\,\kb\,\kc+
\kc^2\right)+\kb\,\kc\,(\kb+\kc)}{\kT^2}\biggr], \\
\cG_2 (\vka,\vkb,\vkc) 
&=& \frac{H_0\,k_2\,k_3}{\sqrt{8\,\ka^3\,\kb^5\,\kc^5}\,\Mp}
\biggl\{3\,i\,\ka^3\l[{\rm Ei}\l(i\,\kT\,\ee\r)+i\,\pi\r]+{\rm e}^{i\,\kT\,\ee}
\biggl[-\frac{3}{\ee^3} \nn\\
&-& \frac{i\,\kb\,\kc\left[3\,\ka^2\,(\kb+\kc)+\ka\left(3\,\kb^2+8\,\kb\,\kc+
3\,\kc^2\right)+\kb\,\kc\,(\kb+\kc)\right]}{\kT^2} \nn\\
&+& \frac{3\,\left[-\ka^2+\ka\,(\kb+\kc)+\kb\,\kc\right]}{\ee}
- \frac{\ka\,\kb^2\,\kc^2\,\ee}{\kT}+
\frac{3\,i\,\kT}{\ee^2}\biggr]\biggr\}.\nn\\
\end{eqnarray}
\end{subequations}
It should be mentioned that the integrals have been regulated (as is 
usually done in this context) in the $\eta\to-\infty$ limit to arrive 
at the above results.

\par

Let us now turn to the case of the bouncing scenario.
In the first domain, as in the inflationary case, the integrals have to 
be suitably regulated in the first domain (in the $\eta \to -\infty$ limit) 
to arrive at the required results.
The integrals in the second domain do not entail considering any non-trivial 
limits and can be evaluated in a straightforward manner.
Moreover, the results for the integrals in the second domain prove to be 
rather lengthy.
Therefore, in what follows, we shall provide the results only in the first
domain.

\par

In the ${\bar n}=1$ case, we find that the integrals are given by
\begin{subequations}
\begin{eqnarray}
\cG_1 (\vka,\vkb,\vkc) 
&=& \frac{a_0\,k_0\,{\rm e}^{-i\,\alpha\,\kT/k_0}}
{\sqrt{2}\,\alpha^3\,C_{\phi}\,(\ka\, \kb\, \kc)^{3/2}\,\kT^2}\,
\biggl\{3\,i\,\alpha^3\,k_0\,\ka^3\,\kT^2\,{\rm e}^{i\,\alpha\,\kT/k_0}\,
{\rm Ei}\l(-\frac{i\,\kT\,\alpha}{k_0}\r)\nn\\
& &+\,3\,k_0^4\,\kT^2
+3\,i\,\alpha\,k_0^3\,\kT^3 
+3\,\alpha^2\,k_0^2\,\kT^2\,\l[\ka^2-\ka\, (\kb+\kc)-\kb\, \kc\r]\nn\\
& &-\,\alpha^3\,k_0\,\biggl[3\,\pi\,\ka^5\,{\rm e}^{i\,\alpha\,\kT/k_0} 
+ 6\,\pi\,\ka^4\,(\kb+\kc)\,{\rm e}^{i\,\alpha\,\kT/k_0}
+i\,\kb^2\,\kc^2\,(\kb+\kc)\nn\\
& &+ 3\,\pi\,\ka^3\,(\kb+\kc)^2\,{\rm e}^{i\,\alpha\,\kT/k_0} 
+ 3\,i\,\ka^2\,\kb\,\kc\,(\kb+\kc)\nn\\ 
& &+\, i\,\ka\,\kb\,\kc\,\l(3\,\kb^2+8\,\kb\,\kc+3\,\kc^2\r)\biggr]
+ \alpha^4\,\ka\,\kb^2\,\kc^2\,\kT\biggr\},\\
\cG_2 (\vka,\vkb,\vkc) 
&=& \frac{a_0\,k_0\,k_2\,k_3\,{\rm e}^{-i\,\alpha\,\kT/k_0}}{\sqrt{2}\, 
\alpha\,C_{\phi}\,(\ka \kb \kc)^{3/2}\,\kT^2}\,
\biggl\{k_0^2\,\kT^2 +i\,\alpha\,k_0\,
\biggl[\ka^2\,(\kb+\kc) \nn\\
& &+\, \ka\,\left(\kb^2+4\,\kb\,\kc+\kc^2\r)
+\kb\,\kc\,(\kb+\kc)\biggr]-\alpha^2\,\ka\,\kb\,\kc\,\kT\biggr\}.
\end{eqnarray}
\end{subequations}
While, in the ${\bar n}=3/2$ case, they can be obtained to be
\begin{subequations}
\begin{eqnarray}
\cG_1 (\vka,\vkb,\vkc) 
&=& -\frac{3\,a_0\,k_0\,{\rm e}^{-i\,\alpha\,\kT/k_0}}{4\,
\sqrt{2}\,\alpha^5\,C_{\phi}\,\ka^{3/2}\,\kb^{5/2}\,\kc^{5/2}\kT^2}\, 
\biggl\{90\,k_0^6\,\kT^2 + 90\,i\,\alpha\,k_0^5\,\kT^3 \nn\\
& &-\, 15\,i\,\alpha^5\,k_0\,\ka^3\,\kT^2\,\l(\ka^2-\kb^2-\kc^2\r)\,
{\rm e}^{i\,\alpha\,\kT/k_0}\,{\rm Ei}\l(-\frac{i \kT \alpha }{k_0}\r)\nn\\
& &+\, 30\,\alpha^2\,k_0^4\,\kT^2\,\l[\ka^2-3\,\ka\,(\kb+\kc)-\kb^2-
3\,\kb\,\kc-\kc^2\r]
-\, 15\,i\,\alpha^3\,k_0^3\,\kT^2\,\biggl[\ka^3\nn\\
& &-\,2\,\ka^2\,(\kb+\kc)
+2\,\ka\,\l(\kb^2+3\,\kb\,\kc+\kc^2\r)
+2\,\kb\,\kc\,(\kb+\kc)\biggr] \nn\\
& &-\, 3\,\alpha^4\,k_0^2\,\kT^2\,\biggl[5\,\ka^4-5\,\ka^3\,(\kb+\kc)
+10\,\ka^2\,\kb\,\kc-10\,\ka\,\kb\,\kc\,(\kb+\kc)\nn\\
& &-\,4\,\kb^2\,\kc^2\biggr]
+ \alpha^5\,k_0\,\biggl[15\,\pi\,\ka^7\,{\rm e}^{i\,\alpha\,\kT/k_0}
+30\,\pi\,\ka^6\,(\kb+\kc)\,{\rm e}^{i\,\alpha\,\kT/k_0}\nn\\
& &+\,30\,\pi\,ka^5\,\kb\,\kc\,{\rm e}^{i\,\alpha\,\kT/k_0} 
- 30\,\pi\,\ka^4\,\l(\kb^3+\kb^2\,\kc+\kb\,\kc^2+\kc^3\r)\,
{\rm e}^{i\,\alpha\,\kT/k_0}\nn\\
& &-\,15\,\pi\,\ka^3\,(\kb+\kc)^2\,\l(\kb^2+\kc^2\r)\,
{\rm e}^{i\,\alpha\,\kT/k_0}
+ 12\,i\,\ka^2\,\kb^2\,\kc^2\,(\kb+\kc)\nn\\
& &+\,4\,i\,\ka\,\kb^2\,\kc^2\,
\l(3\,\kb^2+7\,\kb\,\kc+3\,\kc^2\r)
+2\,i\,\kb^3\,\kc^3\,(\kb+\kc)\biggr] 
- 2\,\alpha^6\,\ka\,\kb^3\,\kc^3\,\kT\biggr\}, \nn\\ \\
\cG_2 (\vka,\vkb,\vkc) 
&=& -\frac{3\,a_0\,k_0\,k_2\,k_3\,
{\rm e}^{-i\,\alpha\,\kT/k_0}}{2\,\sqrt{2}\,\alpha^3\,C_{\phi}\,\ka^{3/2}\,
\kb^{5/2}\,\kc^{5/2}\,\kT^2}\,
\biggl\{3\,i\,\alpha^3\,k_0\,\ka^3\,\kT^2\,{\rm e}^{i\,\alpha\,\kT/k_0}\,
{\rm Ei}\l(-\frac{i\,\kT\,\alpha}{k_0}\r) \nn\\
& &+\, 3\,k_0^4\,\kT^2 + 3\,i\,\alpha\,k_0^3\,\kT^3
+3\,\alpha^2\,k_0^2\,\kT^2\,\l[\ka^2-\ka\,(\kb+\kc)-\kb\,\kc\r]\nn\\
& &-\, \alpha^3\,k_0\,\biggl[3\,\pi\,\ka^5\,{\rm e}^{i\,\alpha\,\kT/k_0}
+6\,\pi\,\ka^4\,(\kb+\kc)\,{\rm e}^{i\,\alpha\,\kT/k_0}\nn\\
& &+\,3\,\pi\,\ka^3\,(\kb+\kc)^2\,{\rm e}^{i\,\alpha\,\kT/k_0}
+ 3\,i\,\ka^2\,\kb\,\kc\,(\kb+\kc)\nn\\
& &+\,i\,\ka\,\kb\,\kc\,\l(3\kb^2+8\,\kb\,\kc
+3\,\kc^2\right)+i\,\kb^2\,\kc^2\,(\kb+\kc)\biggr]
+\alpha^4\,\ka\,\kb^2\kc^2\,\kT\biggr\}.\nn\\
\end{eqnarray}
\end{subequations}

\bibliographystyle{JHEP}
\bibliography{ccbsmg-bu-january-2019}

\end{document}